\documentclass{aa}
\usepackage{graphicx}
\usepackage{subfig}
\usepackage{amsmath}    % Advanced maths commands
\usepackage[T1]{fontenc}
\usepackage{ae,aecompl}
\usepackage[english]{babel}
\usepackage{url}
\usepackage{soul}
\usepackage{orcidlink}
\usepackage{siunitx}
\usepackage{graphicx}
\usepackage{fancyhdr}

\usepackage{amssymb}    % Extra maths symbols
\usepackage{adjustbox}
\usepackage{threeparttable} %Footnote in tables
\usepackage{multirow} % for tables
\usepackage{makecell} % for tables
\usepackage[normalem]{ulem}
\usepackage{colortbl} % Table colors
\usepackage[switch]{lineno}

\usepackage{booktabs}
\DeclareMathAlphabet\mathbfcal{OMS}{cmsy}{b}{n}
\definecolor{Gray}{gray}{0.85}
\newcolumntype{a}{>{\columncolor{Gray}}c}

\newcommand{\beq}[1]{\begin{equation}\label{#1}}
\newcommand{\eeq}{\end{equation}}
\newcommand{\sub}[1]{_{\rm #1}}
\newcommand{\beqn}{\begin{equation}}
\newcommand{\eeqn}{\end{equation}}

\newcommand{\rsat}{R_\mathrm{s}}

\newcommand{\roche}{r\sub{R}}
\newcommand{\hill}{r\sub{H}}

\modulolinenumbers[5]
\usepackage[section]{placeins}

\usepackage[todonotes={textsize=footnotesize}]{changes} %comment if you don't want changes
\definechangesauthor[name=Jorge,color=red]{JZ}

\usepackage{color}
\newcommand{\vv}[1]{\textcolor{black}{ #1}}
\newcommand{\new}[1]{\textcolor{black}{ #1}}

\usepackage{natbib,twoopt}
\bibpunct{(}{)}{;}{a}{}{,}
\usepackage[hyphenbreaks]{breakurl}
\hypersetup{
  colorlinks,
  citecolor=cyan,
  linkcolor=magenta,
  urlcolor=teal,
}

\begin{document}

%Proposed titles
%\title{Could Solar System moons have rings?}
%\title{Could Solar System moons have \added[id=JZ]{had} rings?}
%\title{Have the Solar System moons had rings in the past?}
%\title{Rings around Solar System moons: past and future}
\title{The missing rings around Solar System moons}

\author{
    Mario Sucerquia\inst{1,2,3}\fnmsep\thanks{E-mail: \href{mailto:mario.sucerquia@univ-grenoble-alpes.fr}{mario.sucerquia@univ-grenoble-alpes.fr}} \orcidlink{0000-0002-8065-4199}
    \and Jaime A. Alvarado-Montes\inst{4,5} \orcidlink{0000-0003-0353-9741}
    \and Jorge I. Zuluaga\inst{6} \orcidlink{0000-0002-6140-3116}\and \\
    Nicolás Cuello\inst{2} \orcidlink{0000-0003-3713-8073}
    \and Jorge Cuadra\inst{1,3} \orcidlink{0000-0003-1965-3346}
    \and Matías Montesinos\inst{7,3} \orcidlink{0000-0001-9789-5098}
}

\authorrunning{M. Sucerquia et al.}

\institute{
    Departamento de Ciencias, Facultad de Artes Liberales, Universidad Adolfo Ibáñez, Av. Padre Hurtado 750, Viña del Mar, Chile. \label{inst_uai}
    \and Univ. Grenoble Alpes, CNRS, IPAG, 38000 Grenoble, France. \label{inst_IPAG}
    \and Núcleo Milenio Formación Planetaria - NPF, Chile. \label{inst_npf}
    \and School of Mathematical and Physical Sciences, Macquarie University, Balaclava Road, North Ryde, NSW 2109, Australia. \label{inst_Mac}
    \and The Macquarie University Astrophysics and Space Technologies Research Centre, Macquarie University, Balaclava Road, North Ryde, NSW 2109, Australia. \label{inst_Mac2}
    \and SEAP/FACom, Instituto de Física - FCEN, Universidad de Antioquia, Calle 70 No. 52-21, Medellín, Colombia. \label{inst_udea}
    \and Departamento de Física, Universidad Técnica Federico Santa María, Avenida España 1680, Valparaíso, Chile \label{inst_usm}
}
% These dates will be filled out by the publisher
\date{\today}

% Abstract of the paper
\abstract
%Context
{Rings are complex structures that surround various bodies within the Solar System such as giant planets and certain minor bodies. While some formation mechanisms could also potentially foster their existence around (regular or irregular) satellites, none of these bodies currently bear these structures.}
%Aims
{We aim to understand the underlying mechanisms that govern the potential formation, stability, and/or decay of hypothetical circumsatellital rings (CSRs), orbiting the largest moons in the Solar System. This extends to the exploration of short-term morphological features within these rings, providing insights into the ring survival time-scales and the interactions that drive their evolution.}
%Methods
{To conduct this study, we use numerical N-body simulations under the perturbing influence of the host planet and other moon companions.}
%Results
{We found that, as suspected, moons with a lower Roche--to--Hill radius can preserve their rings over extended periods. Moreover, the gravitational environment in which these rings are immersed influences the system's morphological evolution (e.g, ring size), inducing gaps through the excitation of eccentricity and inclination of constituent particles. Specifically, our results show that Iapetus' and Rhea's rings experience minimal variations in their orbital parameters, enhancing their long-term stability. This agrees with the hypothesis that some of the features of Iapetus and Rhea were produced by ancient ring systems, for example, the huge ridge in Iapetus equator as a result of a decaying ring.}
%Conclusion
{From a dynamical perspective, we found that there are no mechanisms that preclude the existence of CSRs and we attribute their current absence to non-gravitational phenomena. Effects such as stellar radiation, magnetic fields, and the influence of magnetospheric plasma can significantly impact the dynamics of constituent particles and trigger their decay. This highlights the importance of future studies on these effects.}

% Select between one and six entries from the list of approved keywords.
% Don't make up new ones.

\titlerunning{Solar system cronomoons}

\authorrunning{Sucerquia et al.}
\maketitle
% -----------------------
\section{Introduction}
\label{sec:intro}

Planetary rings are fascinating features that have captured the interest of researchers for centuries, since the advent of the telescope and the first observations of Saturn's rings by Galileo Galilei in 1610. These structures are made up of many small particles, ranging in size from micrometres to several metres, and their composition can also vary depending on the characteristics of the planet, such as its size, gravity and magnetic field, and the environment surrounding these systems (for a complete review see \citealt{Tiscareno2013}). 

Planetary rings have been extensively studied by both remote and in-situ observations. In recent decades, flyby encounters by spacecraft such as Pioneer, Voyager, New Horizons, and other missions that visited the giant planets have revealed a wealth of fascinating features in these objects (e.g., \citealt{Tiscareno2013, Sicardy2018, Lauer2018, Tiscareno2018}), such as Saturn's Cassini Division, density waves caused by resonances with nearby moons, spokes, arcs ringlets and propeller-shaped features, among others (see e.g. \citealt{Charnoz2018} and references therein).

The characteristics found within rings are largely acknowledged to arise from the intricate interplay between individual ring particles and their surrounding environment. This intricate environment encompasses the gravitational influence of the host planet, the proximity of other satellites situated within or near the rings, the influx of interplanetary matter that interacts with the rings, collision events, radiation pressure, electrostatic forces, photophoresis, cosmic rays, and others. Also, rings are continuously replenished through material emanating from neighbouring satellites. Hence, comprehending the dynamics and behaviour inherent to these intricate systems necessitates a multidisciplinary methodology, which amalgamates observational data, theoretical models, and numerical simulations. 

Rings have also been observed encircling various small objects, including the centaurs Chariklo and Chiron \citep{Braga-Ribas2014, Ortiz2015, Wood2017, Berard2017}, as well as the trans-Neptunian dwarf planets Haumea and Quaoar \citep{Ortiz2017, Morgado2023, Sicardy2022}, through stellar occultations. Nonetheless, neither the rocky planets of the inner Solar System nor the moons of any planet possess planetary rings. This is intriguing, given that rings should easily form in the Solar System due to close encounters with comets and asteroids, or grazing collisions on their surfaces, which could potentially eject substantial amounts of material into orbit (e.g. \citealt{Charnoz2018}). 

It is important to mention that, on the one hand, satellites have experienced periods of great violence due to impacts from comets, asteroids, and possibly other nearby moons, leading to the formation of high-speed ejecta from their surfaces \citep{Zahnle2003, Kirchoff2010, Rickman2017}. On the other hand, moons such as Io, Europa, Ganymede, Enceladus, and Triton, among others, show evidence of high-energy volcanic or cryovolcanic activity (e.g \citealt{GEISSLER2015}). In both scenarios, the resulting material should be ejected towards the orbits of the satellites, aided by their low-intensity gravitational fields. However, there is currently no evidence for dense circumsatellital rings around moons orbiting close planets.

In \citet{Sucerquia2022}, we recently explored the dynamics of rings around moons for exoplanets that are in close proximity to their host star. This study showed that different paths through moons can give rise to a ring, and also concluded, from numerical simulations, that rings made of either ice or dust could survive long enough to be detected and characterised. In fact, these rings were dubbed ``cronomoons'' because of their resemblance to the famous rings of Saturn ({\em Chronos} in Greek mythology). 

In this paper, we go further and aim to explore, from a dynamical perspective, whether some of the moons of the Solar System can retain circumsatellital rings (CSR for short), taking into account the perturbative effects of the gravitational environment of the host planet and other moons within the system. This work provides some clues explaining the current absence of these features and also revisiting the origins of some satellites in orbit and surface features of certain satellites that can be elucidated by the presence of ancient rings around actual moons. Examples such as the hypothetical rings of Rhea and the equatorial ridge of Iapetus, which will be introduced and discussed in detail later, are pertinent to this discussion. Furthermore, we seek to understand the morphological and dynamical characteristics of these hypothetical structures, and how they would evolve over time.

To provide a background for our study, we briefly describe the properties of Solar System satellites and rings in Sect.~\ref{sec:satellites}. We describe our simulation setup in Sect.~\ref{sec:setup}, while we present the findings of our simulations in Sect.~\ref{sec:results}. Finally, in Sect.~\ref{sec:discussion}, we discuss the implications of our results for the search and characterisation of rings around moons and planets in the Solar System and beyond.

\section{Solar System satellites and rings}
\label{sec:satellites}

%FFFFFFFFFFFFFFFFFFFFFFFFFFFFFFFFFF
%FFFFFFFFFFFFFFFFFFFFFFFFFFFFFFFFFF
\begin{figure*}
	\includegraphics[width=\textwidth]{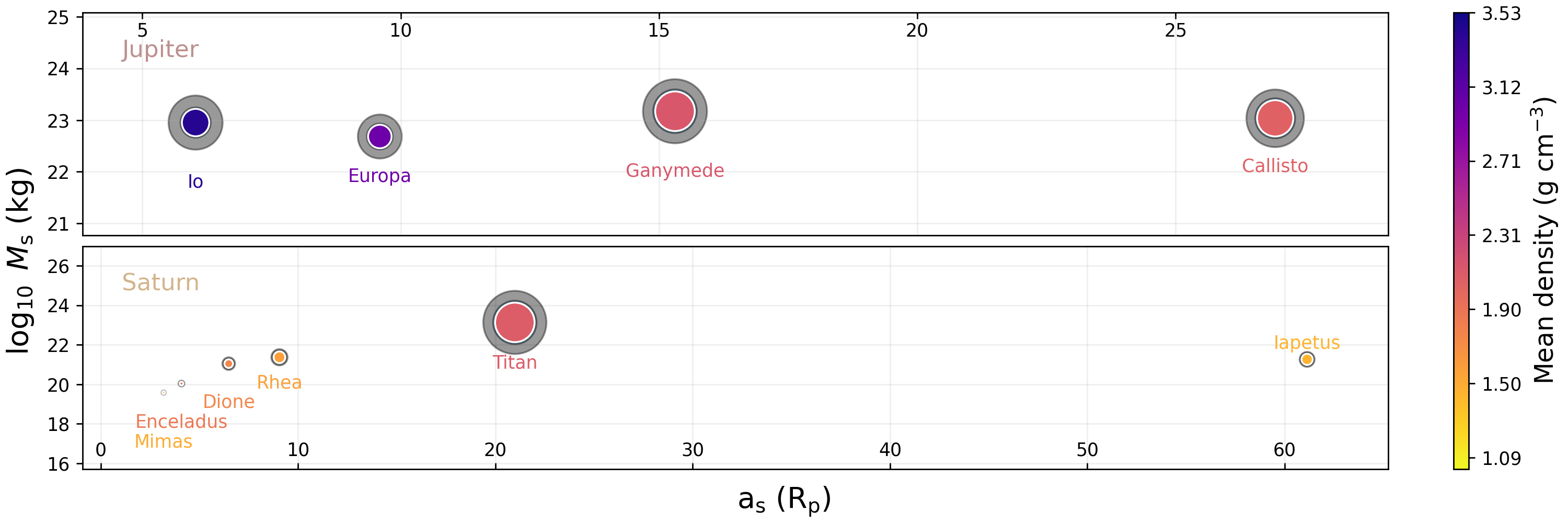}
 	\includegraphics[width=\textwidth]{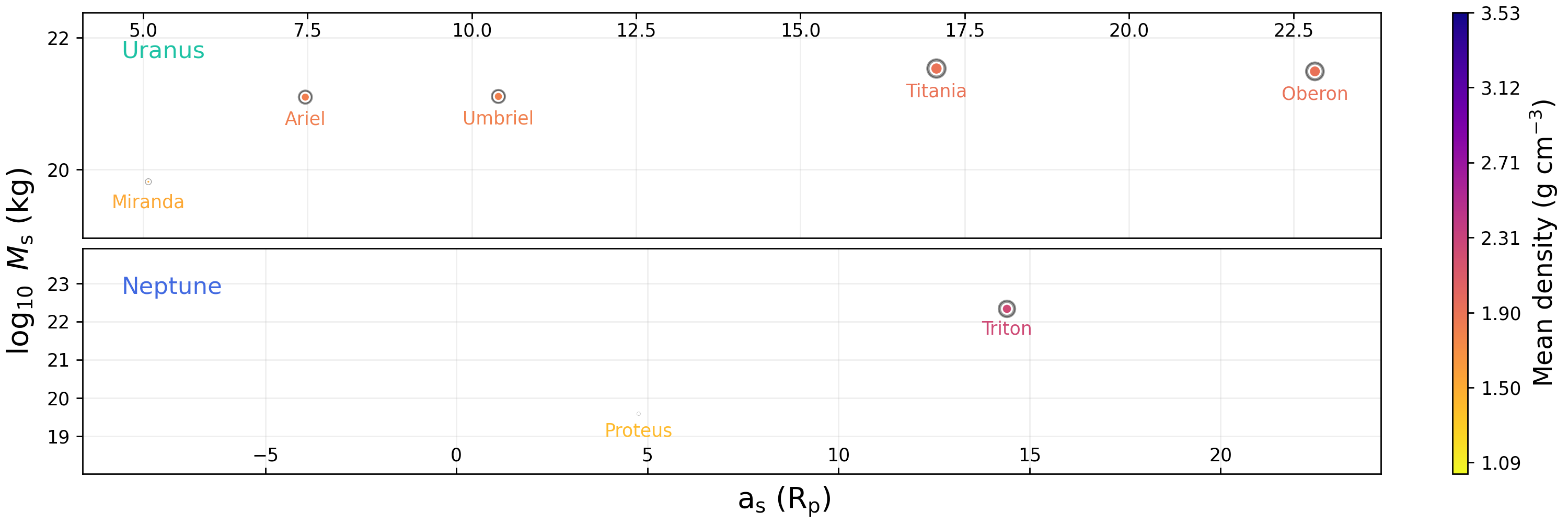}
    \caption{Moon mass versus moon semi-major axis for our sample of large moons around giant planets (two upper panels) and ice giants (two lower panels). The x-axis of each panel is normalised with respect to the radius of each host planet. The radius of the filled circles represents moon size and the shaded area around them shows the size of the ring system (as defined in Sect.~\ref{sec:setup}). Note that moon and rings size are scaled using Titan as a standard size. The colour bar represents moon density.}
    \label{fig:satellites-prop}
\end{figure*}
%FFFFFFFFFFFFFFFFFFFFFFFFFFFFFFFFFF
%FFFFFFFFFFFFFFFFFFFFFFFFFFFFFFFFFF

The Solar System has at least 290 moons\footnote{For updates see \url{https://ssd.jpl.nasa.gov/}}, most of which orbit giant planets and a few rocky and dwarf planets. The majority of the moons around giant planets are regular satellites, namely, satellites with direct orbits and close-by orbits, formed in-situ around the planet. However, irregular moons, i.e. moons probably captured from the interplanetary environment, differ from regulars in terms of their physical and orbital properties, such as an exotic composition or retrograde spin and orbital motion \citep{Hall2016}. Irregular moons do not constitute long-lived isolated objects, but rather rapidly evolve and decay due to gravitational and tidal interactions with the host planets and other sister moons (see, for example \citealt{Sucerquia2019,Sucerquia2020}, and references therein). These moons can also undergo constant collisions that transform their surface and form circumplanetary ring systems from the remnants of a tidal break-up.

Jupiter and Saturn have the largest number of moons in the Solar System. According to the moon classification  by \citet{Hall2016}, Jupiter is currently orbited by 4 large moons, 12 small moons, and 51 very small moons, while Saturn has 1 large moon, 4 medium-sized moons, 21 small moons, and 36 very small moons\footnote{These numbers could change after new discoveries but also if the definition of ``moon'' changes in the future.}. The Uranus system is the smallest in terms of mass, with 27 moons, 4 are medium-sized moons, and 23 are small moons (no large, regular moons orbit the planet). Neptune has the smallest number of moons, with a total of 14 moons, of which only one, Triton, is classified as a large moon, although its origin is still debated. Fig. \ref{fig:satellites-prop} shows the main properties of the large and medium-sized moons of the giant planets (Jupiter and Saturn) and the ice giants (Uranus and Neptune), including their normalised position to the radius of the host planet. The moon size as well as the size of their ring system is normalised using Titan as a standard (Refer to Sect.~\ref{sec:setup} for their definitions).

The rings of the giant planets have a wide variety of features, some of which are common and some of which are quite different. For example, most of them are confined to between 2 and 3 times the radius of the planet and have a small thickness compared to their diameter. However, their composition, origin, and morphological characteristics can vary from one system to another. For instance, Saturn's rings consist of countless small particles, mostly made of water ice with some rocky material mixed in. There are various major divisions within Saturn's rings, including the Cassini Division, the Encke, and the Keeler gaps. The rings are continuously shaped by several small moons, known as ``shepherd moons'', which exert gravitational forces on the particles, shaping and structuring the rings. Particles within the rings can create waves and wakes as they interact with each other and with the nearby moons. Also, the rings of Saturn exhibit mysterious radial features known as ``spokes'' which are sometimes visible and thought to be the result of electrostatic forces \citep{Hartquist2003}.

In contrast, Jupiter's rings are made up of small particles of icy and rocky dust. The majority of these particles are tiny, with sizes ranging from micrometers to a few centimeters. These rings are much thinner and less massive than those of Saturn and have a relatively narrow width of only a few thousand kilometres. The density of material in Jupiter's rings is very low, with only a few particles per cubic centimeter. Additionally, small ``shepherding'' moons, such as Metis and Adrastea, confine the rings and maintain their narrow orbits, similar to Prometheus and Pandora shepherding Saturn's F ring \cite{Hall2016}

Uranus' rings are made up of small particles of ice and dust, similar to those of Jupiter. Also, like Saturn and Jupiter, Uranus has small moons that help to shape and maintain the structure of the rings, such vertical high scale is relatively narrow, and are divided into several distinct regions with different densities and particle sizes \citep{Porco1987,Goldreich1987}. Uranus's rings are thought to be relatively young, estimated to be only a few hundred million years old \citep{Esposito_2002, Esposito1989}.

\new{Following \cite{DePater2018}}, the rings of Neptune differ from those of other giant planets in a few key ways. Composed primarily of dust particles and small rocks, with some water ice, these rings (e.g., Adams and Le Verrier Rings) are maintained by small moons that help shape their structure \new{(e.g., Galatea and Despina, respectively)}. Notably, Neptune's rings contain bright arcs or segments that are believed to be the product of gravitational resonances with nearby moons. Additionally, the rings are divided into several distinct regions, each with varying densities and particle sizes.

There is a close relationship between the satellites of \new{the} giant planets and their rings. Satellites may be destroyed by the planet's tidal forces to form massive rings, and, due to the gravitational viscosity (the process where particles within the rings interact through their gravitational fields, leading to angular momentum and energy transfer) of very old rings, they may end up forming small satellites with portions of their material \citep{Charnoz2010, Cuk2020}.  Also, some authors have postulated that Jupiter's tenuous ring system may have been formed from material ejected from the moons Amalthea, Thebe, Metis, Adrastea, and other parent bodies during high-velocity impacts (see, e.g. \citealt{Burns2004}). Meanwhile, Saturn's F ring \new{may have} formed when Prometheus and Pandora collided with each other and were partially disrupted, and the \new{very active} E ring of Saturn is believed to be maintained from the particles ejected by cryovolcanic activity on the surface of Enceladus and by particles released after continued micro-meteoroid collisions against this moon \citep{Hyodo2015, Spahn2006}.

%F ring may have formed when
%For instance, Quaoar's narrow ring is expected to accrete into a small satellite and disappear over a short time scale, if it is confined to be placed near a 1:3 Spin-Orbit resonance \citep{Sicardy2022}.

Despite the rich diversity of moons within the Solar System, it is intriguing to note the absence of ring systems around them. While many moons are known to encircle giant planets, the phenomenon of moon-hosted rings remains notably absent. This peculiarity prompts a question: Why do moons lack the ring structures so commonly observed around their parent planets? In the subsequent section, we will delve into our approach of employing numerical simulations to explore the long-term stability and morphological attributes of these enigmatic structures.

\section{Simulation setup}
\label{sec:setup}

In this work, the study of the dynamical stability of circumsatellital particles involves the integration of the equations of motion of a set of test particles that evolve under the gravitational influence of the moon, the planet, and the sibling satellites. This approach, for instance,  allows for the constraints of the radii of the inner and outer rings, as well as their stability, lifespan, fate, and morphology. 

In our numerical experiments, we decide to excluded satellites with a mass less than $10^{19}$~kg (on the order of the mass of Mimas), and those with an average density lower than that of water ice to ensure they possess a spherical shape. In other words, we included all major, intermediate, and some selected minor moons from the aforementioned list. To establish the initial conditions of the particles, we used data imported from the NASA Horizons program for the main planet and its moons, at an arbitrary date chosen after the Juno mission\footnote{Specifically, on October 11, 2021, at 3:33 pm.}. \new{The physical and orbital parameters for the bodies involved in each simulation are provided in Table \ref{tab:physels}. The ring particles were initialized in circular orbits ($e=0$) and were coplanar with the moon’s orbit, sharing the same inclination and longitude of the ascending node, as indicated in the aforementioned table. Additionally, we considered the Roche limit ($\roche$) as a guide for the maximum outer limit of the rings, but not necessarily as a restriction.} The Roche limit ensures that circumsatellite particles do not coalesce into larger ones to form new sub-satellites. We assumed that the planet and the satellite are spherical and that the hypothetical rings are coplanar and co-rotating with the satellite's orbit around the planet. The Roche Limit is defined as the orbital limit within which a body's self-attraction (particles) is exceeded by the tidal forces of a primary body (host moon):
\begin{equation}
    \roche = r_\mathrm{m} \left( 2 \frac{\rho_\mathrm{m}}{\rho_\mathrm{part}}\right)^{\frac{1}{3}} \,\,\,,
    \label{eq:roche}
\end{equation}
where $r_\mathrm{m}$ and $ \rho_\mathrm{m}$ are the radius and density of the moon, respectively; and $\rho_\mathrm{part}$ is the density of the CSR, assumed to be equal to that of ice of water (0.917~g/cm$^3$).

Additionally, it is pertinent to elaborate on the so-called Hill radius (both primary and secondary; see Fig.~\ref{fig:satellites-prop}). This radius represents the extent of the gravitational domain of a secondary object in the presence of a primary object, and it also defines the maximum distance at which particles maintain dynamic stability around the secondary object. The Hill radius is defined here following the notation by \citet{Domingos2006}
\begin{equation}
r_{\mathrm {H} }\approx c \; a(1-e){\sqrt[{3}]{\frac {m}{3M}}},
\label{eq:hill}    
\end{equation}
where $a$, $e$, $m$ are the semi-major axis, eccentricity, and mass of the secondary (in this case the satellite), $M$ is the mass of the primary (the host planet), and $c$ is a constant value related to the dynamical stability of the particles: if $c=1$ we recover the traditional expression for the primary Hill radius, if $c \approx 0.4895$ or $c \approx 0.9309$ the expression is the secondary Hill radius, or the stability limit for prograde and retrograde satellites, respectively \citep{Domingos2006}. Beyond this critical threshold, it is widely postulated that particles are subject to the perturbations of evection/eviction resonances \citep{Vaillant2022}, rendering them dynamically unstable on short time-scales.

%FFFFFFFFFFFFFFFFFFFFFFFFFFFFFFFFFF
%FFFFFFFFFFFFFFFFFFFFFFFFFFFFFFFFFF
\begin{figure}
    \includegraphics[width=\columnwidth]{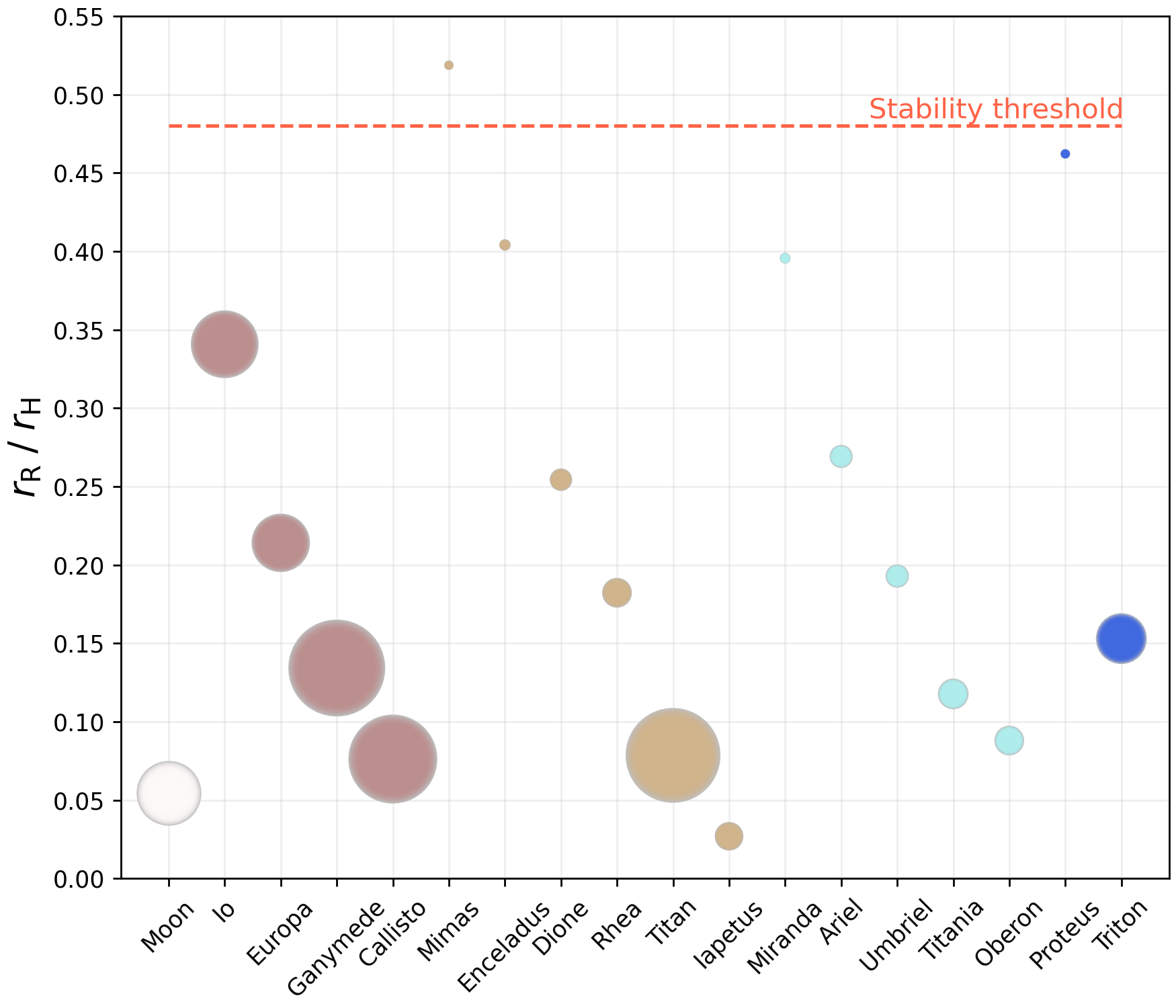}
    \caption{Ratio between the gravitational domain (Hill radius) and the theoretical ring size (Roche limit) for our sample of large moons in the Solar System. Each colour denotes a different host planet. The dashed red horizontal line marks the orbital stability threshold, above which ring particles are likely to remain gravitationally bound.}
    \label{fig:orbdom}
\end{figure}
%FFFFFFFFFFFFFFFFFFFFFFFFFFFFFFFFFF
%FFFFFFFFFFFFFFFFFFFFFFFFFFFFFFFFFF

Therefore, by calculating the ratio of the Roche limit to the Hill limit ($\roche / \hill$) for each of the selected satellites, we can gain insights into the stability of potential sub-satellites (a satellite of a satellite) or rings. For instance, the closer these limits are to each other, the more susceptible the particles become to escaping, colliding with the planet, or colliding with each other due to excitation of their eccentricities.
Fig.~\ref{fig:orbdom} presents a comparison of these dynamical stability boundaries. This can be interpreted as follows: the higher the ratio $\roche / \hill$ for each moon the more unstable the hypothetical rings surrounding it should be. For instance, particles surrounding Io, Mimas, Enceladus, Miranda, and Proteus are likely to be unstable, as the ratio between the Roche limit and the Hill limit is close to 0.49. Particularly in the case of Mimas, where this ratio exceeds 0.5, we predict a complete instability of its rings. Conversely, rings around Iapetus are expected to be the most stable among the set of moons analysed.

Based on the aforementioned assumptions, we established the boundaries of the circumsatellital ring system. We set the ring's inner radius to 1.05 times the satellite radius, $\rsat$, and its outer radius to 1.2 $\roche$. The rationale behind selecting such outer radius stems from the fact that if the rings originated from a collision, the resulting debris could be ejected over long distances\footnote{However, some objects such as Quaoar's rings are not restricted by this limit}. Moreover, to avoid oversampling problems, we randomly generated the initial positions of 1000 particles that compose the ring system, employing blue-noise sampling algorithms. In particular, we use the {\tt fibpy} algorithm based on the work of \cite{VOGEL197}\footnote{\texorpdfstring{\url{https://github.com/matt77hias/fibpy}}{https://github.com/matt77hias/fibpy}}. 
\new{Additionally, to complete our setup, we included the host planet and its other large moons. Furthermore, we did not consider the zonal coefficients \(J_i\) of each planet to reduce computational load. Including these coefficients in future work could improve the description of the gravitational field and thus, the prediction of satellite and ring orbits.} In the case of Earth's moon, we also included the perturbative effect of the Sun due to its relative proximity. Finally, to run our simulations employing the {\tt REBOUND} N-body code \citep{rebound}, with the integration being performed over 1000 years using the symplectic Wisdom-Holman integrator, WHFast \citep{reboundwhfast,wh}. The integration duration corresponds to approximately 450 periods of Iapetus, the moon with the longest orbital period included in our simulations, which is around 79 Earth days. Finally, we used a time-step of the order of a few minutes.

To assess the stability and morphological evolution of the CPRs, we conduct the following experiments:

\begin{enumerate}
    \item \textbf{Moon stability:} We simulate the evolution of each planet's satellites over a span of one million years to verify the regularity of the orbits within the gravitational scheme of our interest.
    
    \item \textbf{Long-term integration:} We study the dynamics of representative ring particles, specifically the innermost and outermost particles of the system. Their orbits are then classified after one million years of evolution within a system composed of the planet and the satellites that meet the above criteria.
    
    \item \textbf{MEGNO analysis:} We compute the MEGNO parameter (an acronym for Mean Exponential Growth factor of Nearby Orbits) in order to test the dynamical structure of the ring particles and evaluate their long-term orbital stability.
    
    \item \textbf{Gravitational Environment:} We study a specific case wherein the evolution of ring morphology is compared for a system where the rings are isolated (i.e., only the planet, satellite, and rings are included) and when the gravitational perturbations of other sibling moons (or the Sun, in the case of Earth's moon) is taken into account.
    
    \item \textbf{Morphology:} We analyse the evolution of the ring system for each satellite in an environment that considers the \vv{perturbation} of the planet and other satellites. This will allow us to evaluate their influence on particles that lead to the dissipation of the rings or the emergence of resonant structures in their orbital elements.
\end{enumerate}

%################################
%#     SECTION
%################################

\section{Results and analysis}
\label{sec:results}

\begin{figure*}
\centering
    \includegraphics[width=0.6\textwidth]{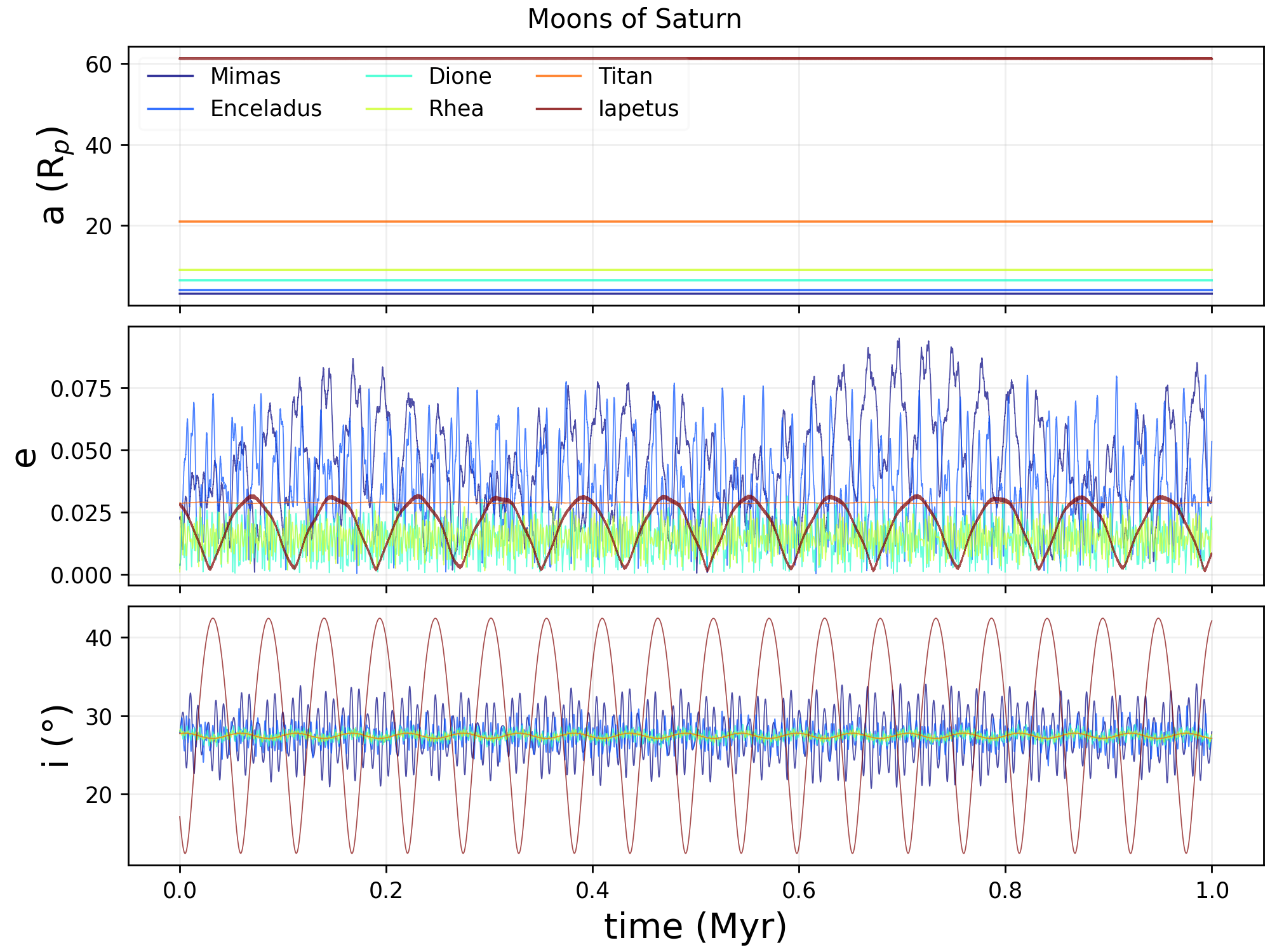}
    \caption{ Orbital characteristic evolution of the moons within Saturn's system over a span of 1 Myr. The three horizontal panels display, from top to bottom, the evolution of the semi-major axis ($a$), eccentricity ($e$), and inclination ($i$) for each moon (respectively).}
    \label{fig:moon_evo}
\end{figure*}
%FFFFFFFFFFFFFFFFFFFFFFFFFFFFFFFFFF
%FFFFFFFFFFFFFFFFFFFFFFFFFFFFFFFFFF

\subsection{Satellite stability}
\label{subs:sat_stab}

%FFFFFFFFFFFFFFFFFFFFFFFFFFFFFFFFFF
%FFFFFFFFFFFFFFFFFFFFFFFFFFFFFFFFFF
\begin{figure}
    \includegraphics[width=\columnwidth]{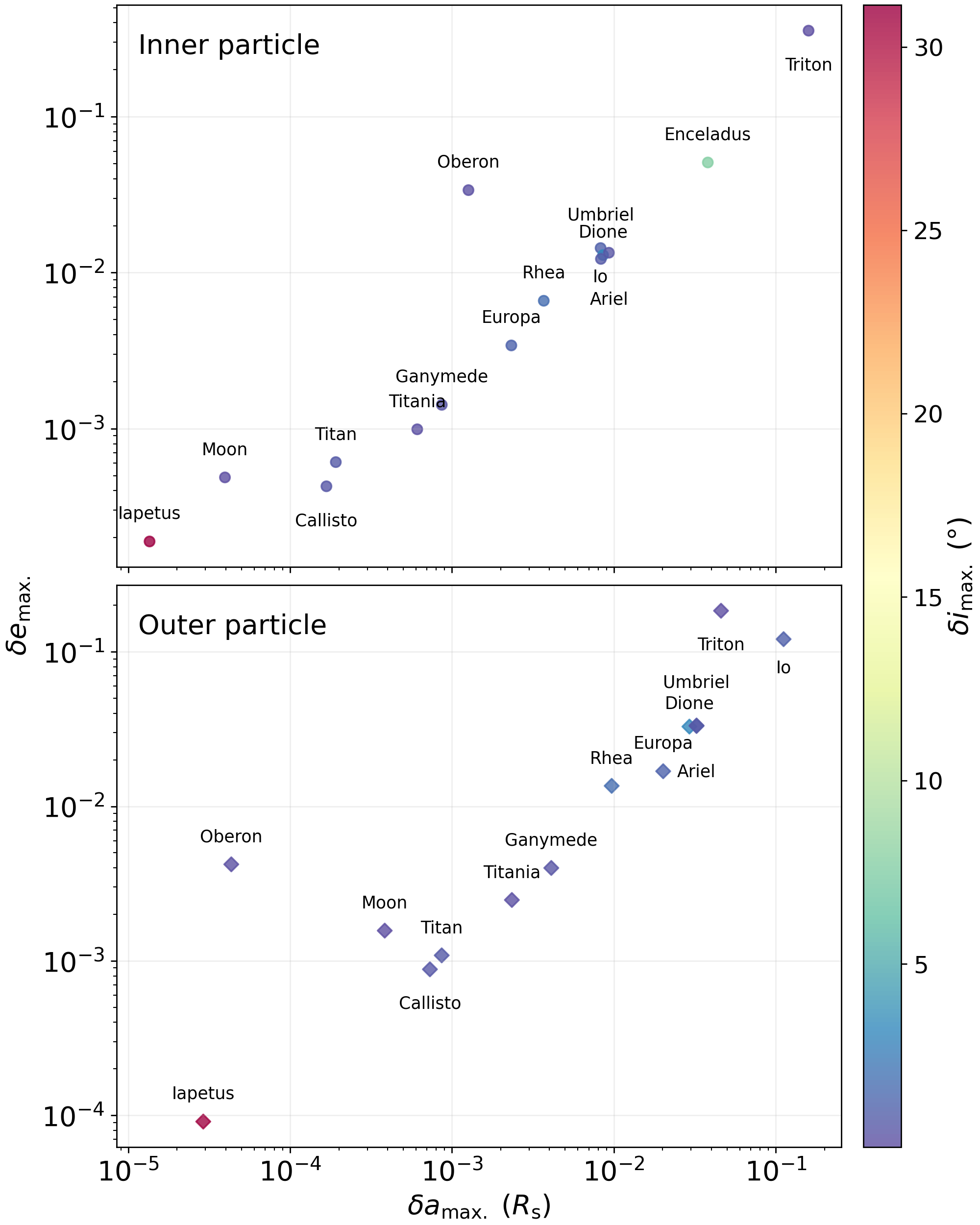}
    \caption{ \new{The upper panel summarises the long-term orbital variations, specifically $\delta a$, $\delta e$, and $\delta i$ (the colourbar), for an isolated inner (upper panel) and outer particle (lower panel) orbiting the moons over a span of 1 Myr. The missing label for Enceladus in the lower panel means that the orbit was unbound.}}
    \label{fig:lt_delta}
\end{figure}
%##############FIGURE##############

Utilising the same initialisation system delineated in the preceding section, we numerically integrated the moon systems of Earth, Jupiter, Saturn, Uranus, and Neptune. This was undertaken to ensure that, in the simulations of the ring particles, the moon system behaved regularly, as anticipated from long time-scale integrations such as those presented by \citealt{Kane_2022}, and references therein. For instance, \new{Fig.~\ref{fig:moon_evo} showcases the orbital evolution of Saturn’s satellites, where no significant changes in the evolution of the semi-major axes are discernible}. Concurrently, the eccentricity and inclination undergo regular periodic variations in their evolution, and there is no presence of secular perturbations that destabilise the system within the evaluated time interval. Saturn was used as an example, but the other systems showed a similar behaviour.
%FFFFFFFFFFFFFFFFFFFFFFFFFFFFFFFFFF
%FFFFFFFFFFFFFFFFFFFFFFFFFFFFFFFFFF
\begin{figure*}
    \includegraphics[width=\columnwidth]{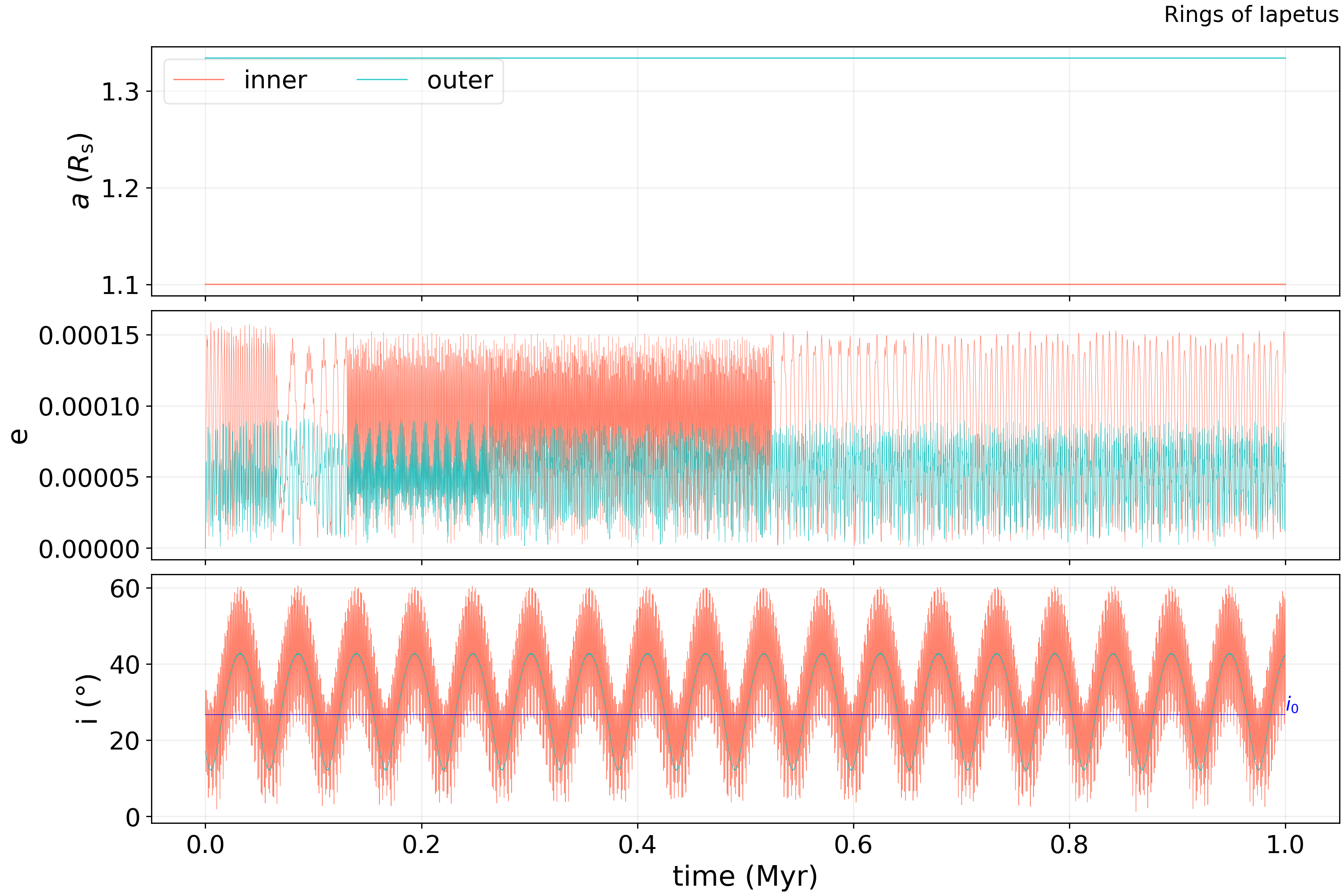}
    \includegraphics[width=\columnwidth]{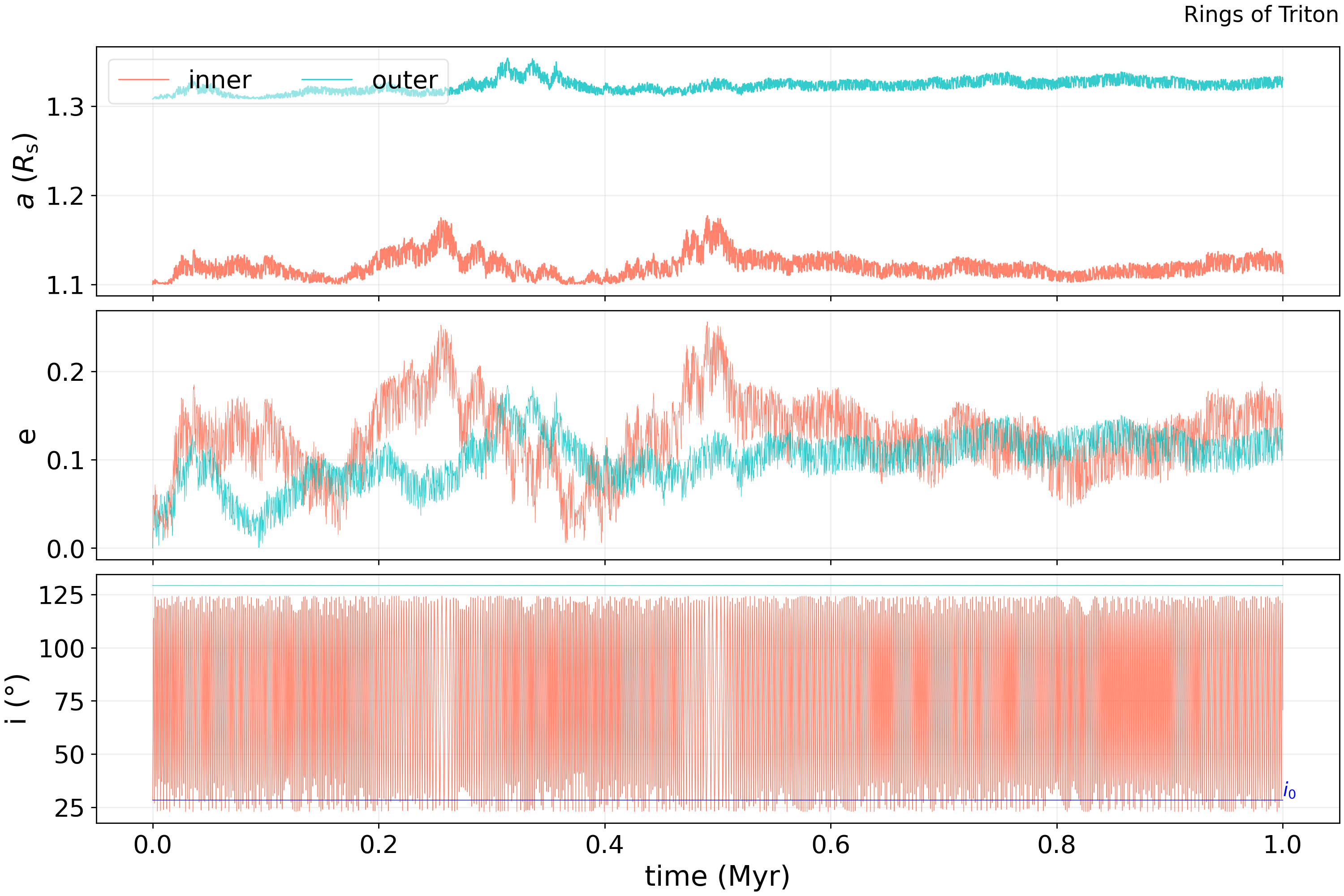}
    \caption{Orbital elements as a function of time for particles orbiting Iapetus (left panels) and Triton (right panels). Iapetus and Triton are the most extreme cases observed in the upper panel of Fig.~\ref{fig:lt_delta}. The horizontal blue line represents the initial inclination ($i_\mathrm{0}$) of the particles. }
    \label{fig:singpart_evol}
\end{figure*}
%##############FIGURE##############

\begin{table}[]
\begin{tabular}{ccc|ccc}
\multicolumn{1}{l}{{\color[HTML]{000000} Satellite}} & {\color[HTML]{000000} Inner} & {\color[HTML]{000000} Outer} & \multicolumn{1}{l}{{\color[HTML]{000000} Satellite}} & {\color[HTML]{000000} Inner} & {\color[HTML]{000000} Outer} \\ \hline
{\color[HTML]{9B9B9B} Moon} & {\color[HTML]{000000} R} & {\color[HTML]{000000} R} & {\color[HTML]{CD9934} Titan} & {\color[HTML]{000000} R} & {\color[HTML]{000000} R} \\
{\color[HTML]{986536} Io} & {\color[HTML]{000000} R} & {\color[HTML]{000000} I} & {\color[HTML]{CD9934} Iapetus} & {\color[HTML]{000000} I} & {\color[HTML]{000000} R} \\
{\color[HTML]{986536} Europa} & {\color[HTML]{000000} R} & {\color[HTML]{000000} I} & {\color[HTML]{34CDF9} Miranda} & {\color[HTML]{000000} I} & {\color[HTML]{000000} I} \\
{\color[HTML]{986536} Ganymede} & {\color[HTML]{000000} R} & {\color[HTML]{000000} R} & {\color[HTML]{34CDF9} Ariel} & {\color[HTML]{000000} R} & {\color[HTML]{000000} R} \\
{\color[HTML]{986536} Callisto} & {\color[HTML]{000000} R} & {\color[HTML]{000000} R} & {\color[HTML]{34CDF9} Umbriel} & {\color[HTML]{000000} R} & {\color[HTML]{000000} R} \\
{\color[HTML]{CD9934} Mimas} & {\color[HTML]{000000} I} & {\color[HTML]{000000} I} & {\color[HTML]{34CDF9} Titania} & {\color[HTML]{000000} R} & {\color[HTML]{000000} R} \\
{\color[HTML]{CD9934} Enceladus} & {\color[HTML]{000000} R} & {\color[HTML]{000000} R} & {\color[HTML]{34CDF9} Oberon} & {\color[HTML]{000000} I} & {\color[HTML]{000000} I} \\
{\color[HTML]{CD9934} Dione} & {\color[HTML]{000000} R} & {\color[HTML]{000000} R} & {\color[HTML]{3531FF} Proteus} & {\color[HTML]{000000} I} & {\color[HTML]{000000} I} \\
{\color[HTML]{CD9934} Rhea} & {\color[HTML]{000000} R} & {\color[HTML]{000000} R} & {\color[HTML]{3531FF} Triton} & {\color[HTML]{000000} I} & {\color[HTML]{000000} I} \\ \hline
\end{tabular}
\caption{Classification of particle orbit behavior within the CSR, categorized as Inner and Outer. The labels ``R'' and ``I'' denote regular and irregular behavior, respectively, as defined in the text. \new{Different colors are used to identify each planetary system.}}
\label{tab:IR}
\end{table}

\subsection{Long-term stability}
\label{subs:long_term}

We conducted a simulation of the orbital dynamics of massless individual ring particles around the Solar System's moons for a duration of one million years to identify potential sources of instability that may result in considerable variations in the primary orbital elements. For the initial conditions, particles positioned at the ring's inner and outer edges of the respective moon were selected as the more representative particles from the ring. We have limited this simulation to only a few particles, considering that their orbital period is relatively short, spanning only a few hours. This implies that integrating over time scales such as millions of years results in a high computational cost (for instance, each simulation took $\sim$3.5 days). Therefore, our approach has been to seek an optimal balance between the duration of the integration and the associated computational costs.

%Firstly, particles surrounding the moons Mimas, Miranda and Proteus escape within relatively short timeframes, on the order of several decades. In contrast, particles around Enceladus also escape, but only after approximately 20,000 years.

In Fig.~\ref{fig:lt_delta}, we present the maximum excitation of the orbital elements of the particles ($\delta a_\mathrm{max.}$, $\delta e_\mathrm{max.}$, and $\delta i_\mathrm{max.}$) at the end of the simulation for the inner (upper panel) and outer particles (lower panel). In general terms, for all systems, at least one of these evaluated characteristics displays substantial variations. For instance, in the case of Iapetus, the particle shows minuscule $\delta a_\mathrm{max.}/R_{\rm s}$ and $\delta e_\mathrm{max.}$ (of order $10^{-5}$ to $10^{-4}$), but a significant secular excitation in its inclination with $\delta i \sim 30$° for both particles, suggesting that the observed change in inclination is the result of a dynamical behaviour shared among the ring particles. Finally, Triton and Io exhibit the most significant changes amongst the evaluated systems. Here, their semi-major axis and eccentricity display erratic and irregular behaviour (see Fig.~\ref{fig:singpart_evol}). The nature of the instabilities observed might be linked to the relationship between the Roche limit and the Hill radius (refer to Fig.~\ref{fig:orbdom}), as well as the gravitational ``stress'' exerted by the surroundings in the cases of Io and Enceladus. In the case of Triton, these instabilities might be attributed to its retrograde translation, considering that the simulation only included the gravitational effects of the distant and relatively minor moon Proteus, as well as the planet itself.

Table \ref{tab:IR} summarises the behaviour of the innermost and outermost particles in the ring system, corresponding to each satellite evaluated in our simulations, based on a visual analysis of the orbital elements associated to each moon. The behaviours of each orbital parameter for these orbits can be categorised into one of two distinct types. Periodic or regular orbits (R) are characterised by their repetitive nature, recurring over fixed intervals of time (see left panels of Fig. \ref{fig:singpart_evol}). \new{On the other hand, irregular orbits (I) denote a state of unpredictability, \vv{perturbed} by the gravitational influence of other objects within the system, such as the parent planet or moons} (see right panels of Fig. \ref{fig:singpart_evol}).

%Secular orbits (S) are marked by their gradual transformations, with changes primarily driven by the gravitational influence exerted by other entities within the system. Lastly,

\subsection{MEGNO}
\label{subs:MEGNO}
The Mean Exponential Growth factor of Nearby Orbits (MEGNO, \citealt{Cincotta2000}) is a computational technique used to distinguish between regular and chaotic behaviour within a dynamical system. Its functionality relies on examining the exponential divergence or convergence of trajectories in phase space over time. In the context of this study, MEGNO is employed to assess the long-term stability of trajectories within the system. MEGNO value approaches $2$ for regular orbits, while for chaotic orbits it yields higher average values, thus providing insights into the dynamical stability of the ring system. We calculated the MEGNO values for 100 constituent particles of the rings, integrating their equation of motion over a period of 1ky years. The constraints on the timespan and the number of particles were chosen to strike a balance between the accuracy of the mean results and computational costs. This approach is particularly necessary because, as detailed below, each individual moon requires 100 independent simulations for 5 or more bodies (planet and moons) over the defined simulation time.

%FFFFFFFFFFFFFFFFFFFFFFFFFFFFFFFFFF
%FFFFFFFFFFFFFFFFFFFFFFFFFFFFFFFFFF
\begin{figure}
    \includegraphics[width=1.0\columnwidth]{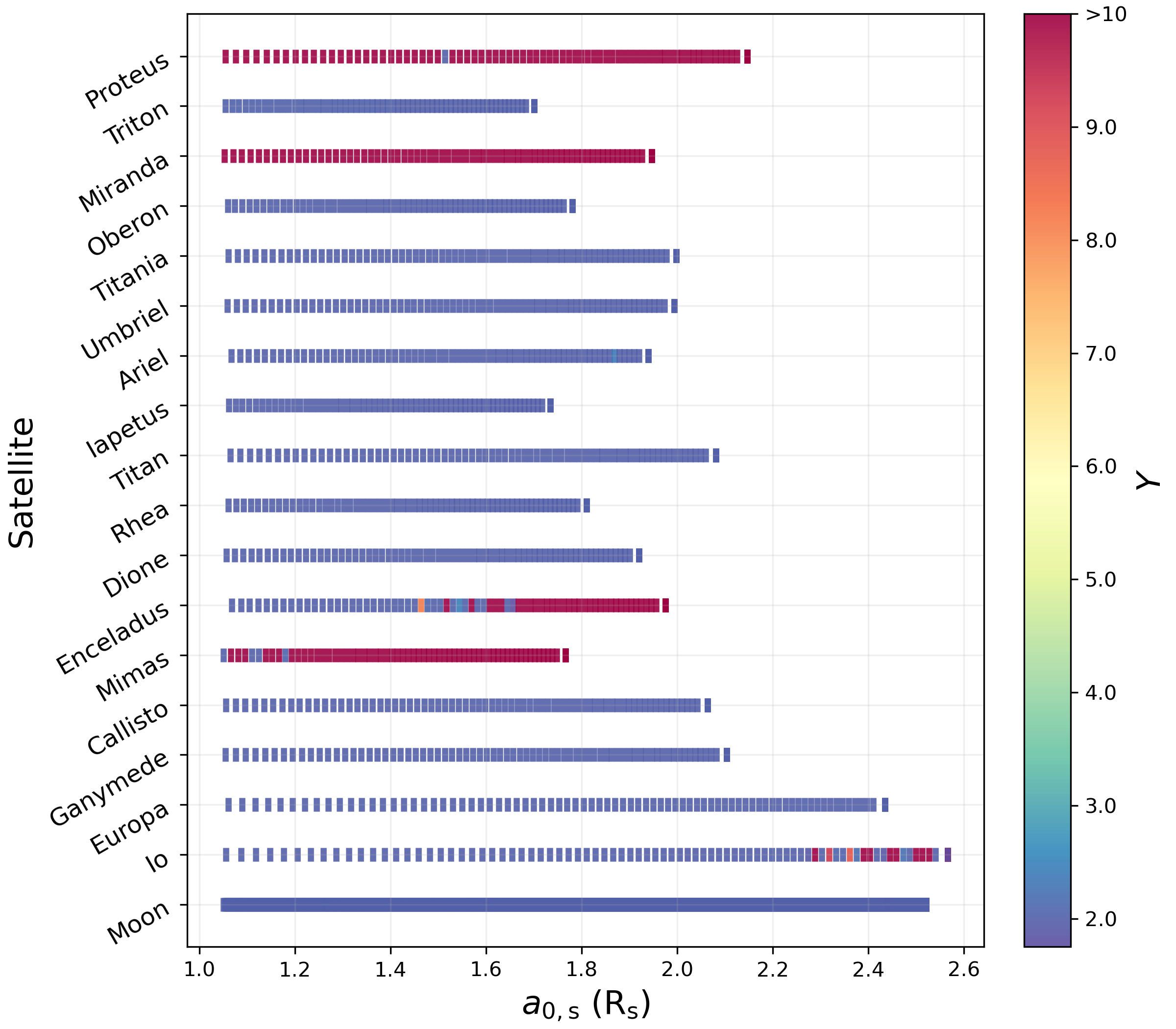}
    \caption{This diagram displays the orbital stability within the ring systems of selected satellites. The horizontal axis represents the initial semi-major axis \( a_\mathrm{0,s} \), in satellite radii \( R_\mathrm{s} \), of the particles in the rings, while the vertical axis lists the various satellites. The MEGNO (\( Y \)) parameter values for each particle are depicted by the colour scale, with shades varying according to the detected degree of orbital stability.}
    \label{fig:megno}
\end{figure}
%FFFFFFFFFFFFFFFFFFFFFFFFFFFFFFFFFF
%FFFFFFFFFFFFFFFFFFFFFFFFFFFFFFFFFF

It should be emphasised that MEGNO evaluates the overall stability of the system. For example, the MEGNO value will deviate from $2$ for the entire system if even a single particle within the ring becomes unstable, which can create an illusion of irregularity. The need to evaluate this parameter for each individual particle within the ring leads to the following sequence of steps. Firstly, we establish the arrangement of the planets, moons and rings according to Sect.~\ref{sec:setup}. Then, we select a ring particle, allow the system to evolve over a predetermined period of time, and compute the relevant parameter. We recalculate the orbit using another particle with a different set of parameters until covering all the particles in the ring. This methodology not only verifies our numerical parameters but also provides a comprehensive understanding of the temporal progress of these ring systems. A summary of the implementation of this approach, using the {\tt REBOUND} software, is depicted in Fig.~\ref{fig:megno}.

In this regard, the prevalence of MEGNO values near 2 suggests the existence of relatively stable orbits, indicating a coherent and predictable motion for a significant portion of particles. In contrast, striking spikes in MEGNO values indicate localised instability and hint at the potential for chaotic behaviour in these regions. Fig.~\ref{fig:megno} substantially corroborates the instability of the rings around the satellites Mimas, Miranda, and Proteus. This result indicates that only a few sharp ringlets could orbit Mimas and Proteus. Regarding Enceladus and Io, the MEGNO analysis reveals instability in the outermost regions of their orbiting rings, thus constraining their width. MEGNO suggests the potential stability of the remaining moons over the evaluated period. However, it is only by analysing the behaviour of semi-major axes, eccentricities, and inclinations that a more comprehensive understanding of the structural dynamics of these systems can be achieved (see Sect. \ref{subs:Morphology}).

It is worth noting that the application of the MEGNO technique, as described in this study, does not consider any interactions between ring particles. These interactions, not studied in this work, may add either instability or stability and are discussed in later sections.

\subsection{The effect of the neighbourhood}
\label{subs:environment}

\citealt{Sucerquia2022} examined the dynamics of rings encircling isolated moons. This scenario applies for Earth but not for other planets in the Solar Systems where moons are typically accompanied by other comparably-sized objects. The perturbations of such bodies can significantly influence the evolution of ring systems.

The above highlights the importance of studying the impact of environmental conditions on ring systems. To this end, we undertook dynamic experiments using the complete assembly of ring particles to assess and compare the system's collective behaviour under two distinct conditions. The first scenario assumes the ring system to be fully isolated from perturbations triggered by neighbouring satellites, whereas the second scenario considers the ring system to be situated within the gravitational field of these satellites. Both scenarios were examined over a period of 1 kyr and included the presence of the host planet.

%FFFFFFFFFFFFFFFFFFFFFFFFFFFFFFFFFF
%FFFFFFFFFFFFFFFFFFFFFFFFFFFFFFFFFF
\begin{figure}
    \includegraphics[width=\columnwidth]{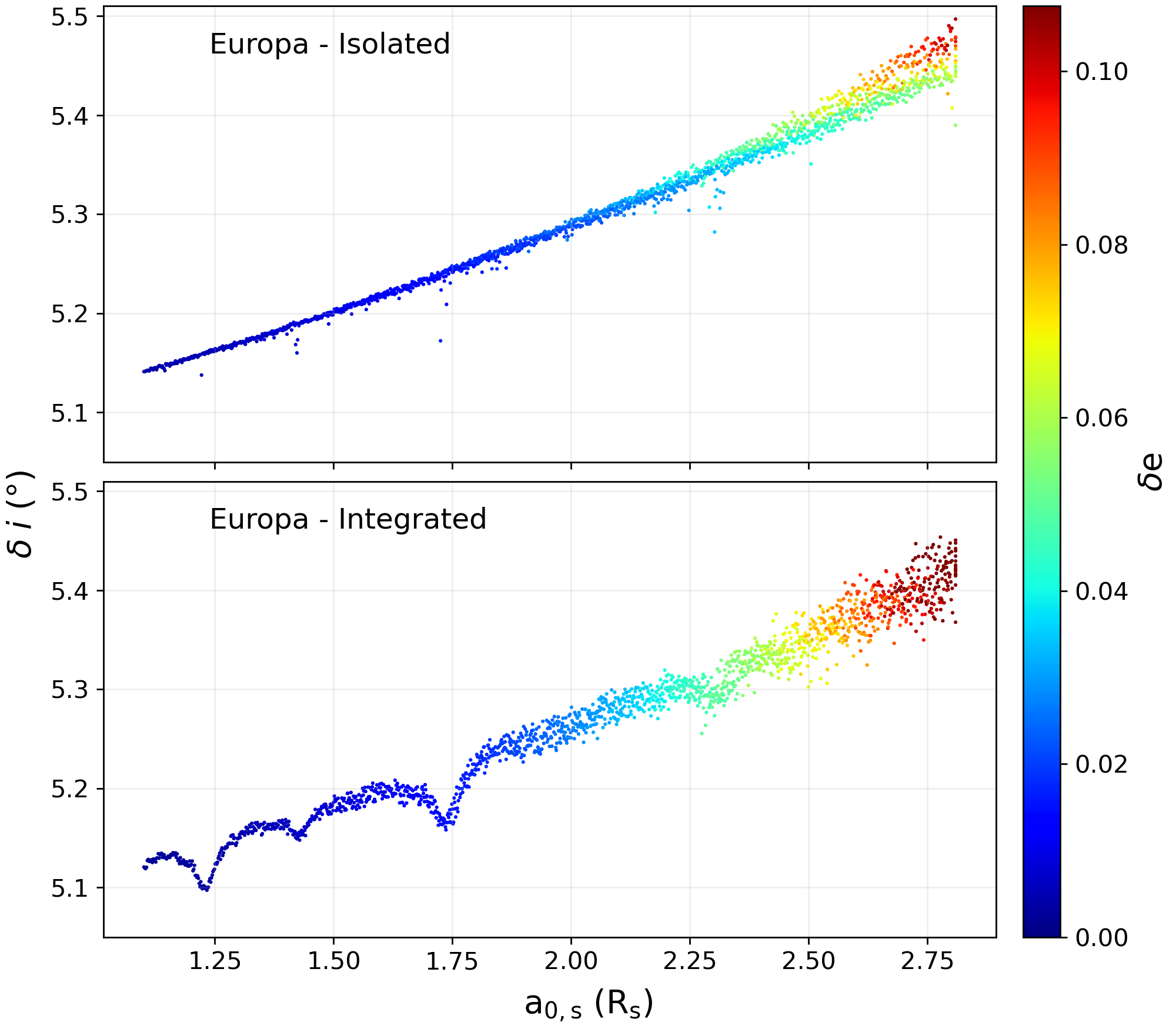}
    \caption{\new{Comparison of Europa's ring particles' orbital elements $\delta i$ and $\delta e$ as a function of their initial semi-major axis $a_{\mathrm{0,s}}$, in the presence (``integrated'') and absence (``isolated'') of other Galilean satellites. This comparison reveals how the presence of these satellites influences the distribution and behavior of the particles.}}
    \label{fig:euro-comp}
\end{figure}

As our study case, we chose Europa as the primary ringed satellite and devised two contrasting test scenarios. The ``Isolated'' scenario refers only to the Jupiter-Europa system and the ``Integrated'' scenario incorporates also its sibling moons: Io, Ganymede, and Callisto. For this experiment, the rings do not lie on the orbital plane of the moon to seek an enhancement of the perturbation effects.

Fig. \ref{fig:euro-comp} depicts the outcome of the system's evolution, clearly illustrating the effects of the environment on the orbital evolution of particles. In the top panel (``isolated''), the absence of resonant interactions among Europa's complex ring system leads to a lack of structure in specific regions, maintaining a more uniform morphology of the ring particles over time. In contrast, the bottom panel (``Integrated'') shows how these resonant interactions generate noticeable structures that alter the morphology of the rings. In this same figure, the colour of the dots indicates changes in particle eccentricity, which are more intense in the integrated case and affect mostly the ring's outer particles. Such perturbations may lead to the loss of particles through collisions, constraining further the size of the ring and even their lifespan. However, a detailed investigation of this phenomenon is outside the scope of this study, as so is analysing individually the effects of companion moons on these hypothetical rings: it is virtually impossible to isolate these moons. In the following sections, we will present the results obtained for the remaining moons.

\subsection{Evolution of Ring Morphology}
\label{subs:Morphology}

%FFFFFFFFFFFFFFFFFFFFFFFFFFFFFFFFFF
%FFFFFFFFFFFFFFFFFFFFFFFFFFFFFFFFFF

%FFFFFFFFFFFFFFFFFFFFFFFFFFFFFFFFFF
%FFFFFFFFFFFFFFFFFFFFFFFFFFFFFFFFFF
\begin{figure}
    \includegraphics[width=\columnwidth]{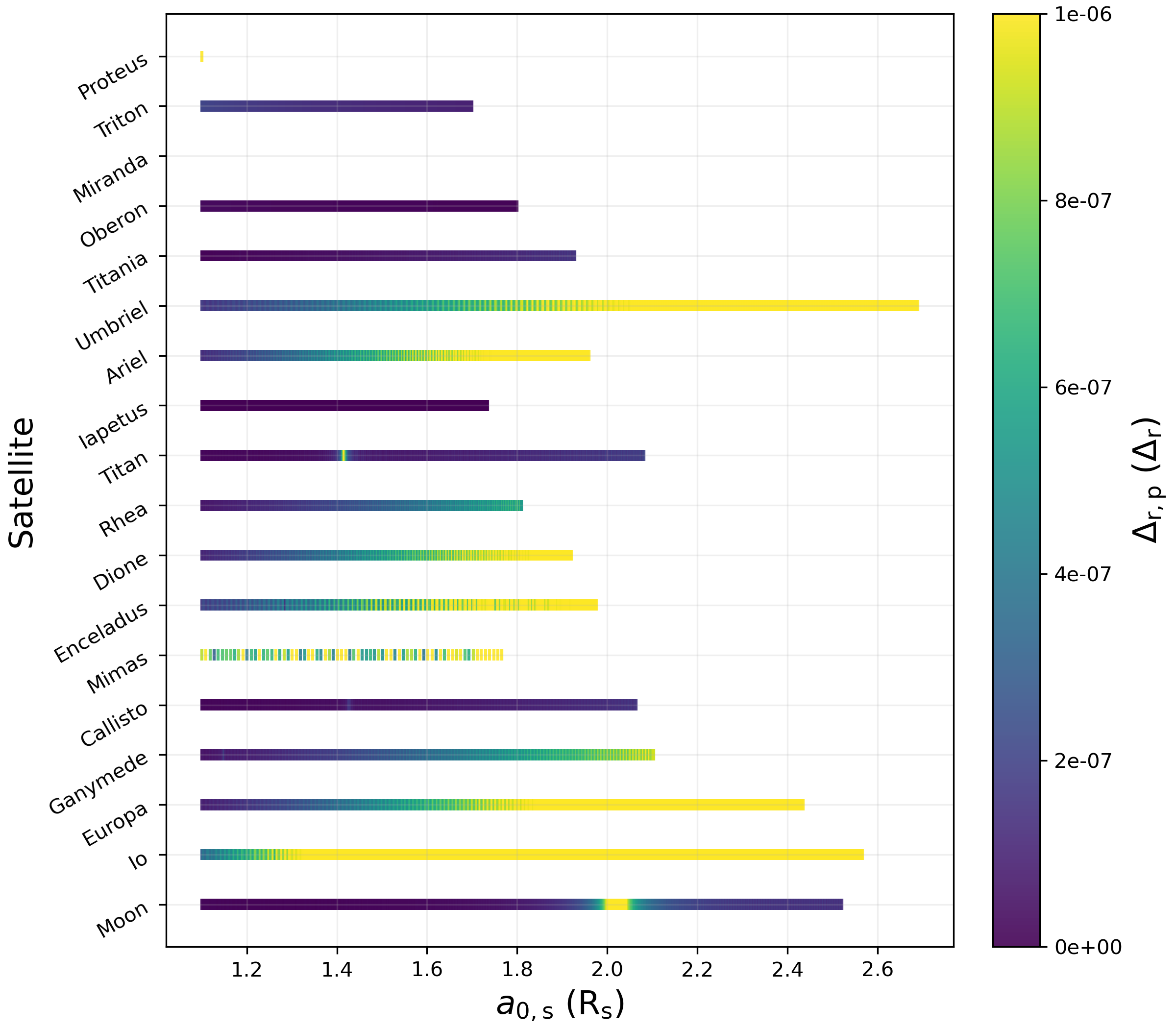}
    \caption{An exhaustive depiction of the orbital behaviour of particles within the ring systems of the investigated satellites. The horizontal axis signifies the initial semi-major axis \( a_\mathrm{0} \), in terms of the satellite radii \( R_\mathrm{0} \), for each constituent particle of the ring. The colour map illustrates the corresponding maximum orbital widening \( \Delta r_\mathrm{p} \) of each ring particle, expressed as a fraction of the initial ring width, upon the completion of the simulations. }
    \label{fig:dQq}
\end{figure}
%FFFFFFFFFFFFFFFFFFFFFFFFFFFFFFFFFF
%FFFFFFFFFFFFFFFFFFFFFFFFFFFFFFFFFF
%FFFFFFFFFFFFFFFFFFFFFFFFFFFFFFFFFF
%FFFFFFFFFFFFFFFFFFFFFFFFFFFFFFFFFF
\begin{figure*}
    \includegraphics[width=1.0\textwidth]{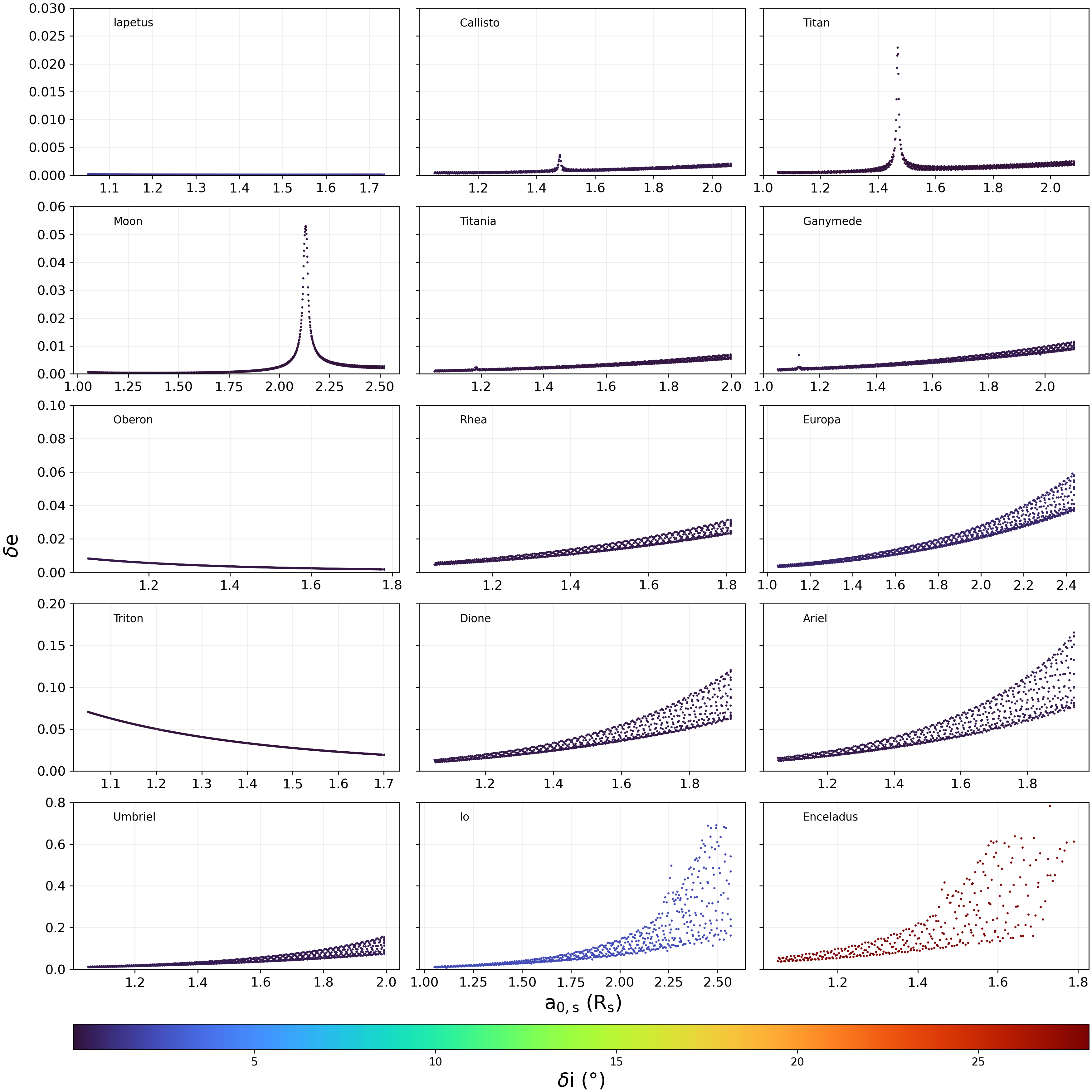}
    \caption{\new{Numerical simulation outcomes for rings around moons. $\delta e$ and $\delta i$ represent the maximum amplitude of the excitation of the ring particles' eccentricity and tilt, respectively. }}
    \label{fig:resonances-e-giant}
\end{figure*}
%FFFFFFFFFFFFFFFFFFFFFFFFFFFFFFFFFF
%FFFFFFFFFFFFFFFFFFFFFFFFFFFFFFFFFF
%FFFFFFFFFFFFFFFFFFFFFFFFFFFFFFFFFF
\begin{figure*}
    \includegraphics[width=1.0\textwidth]{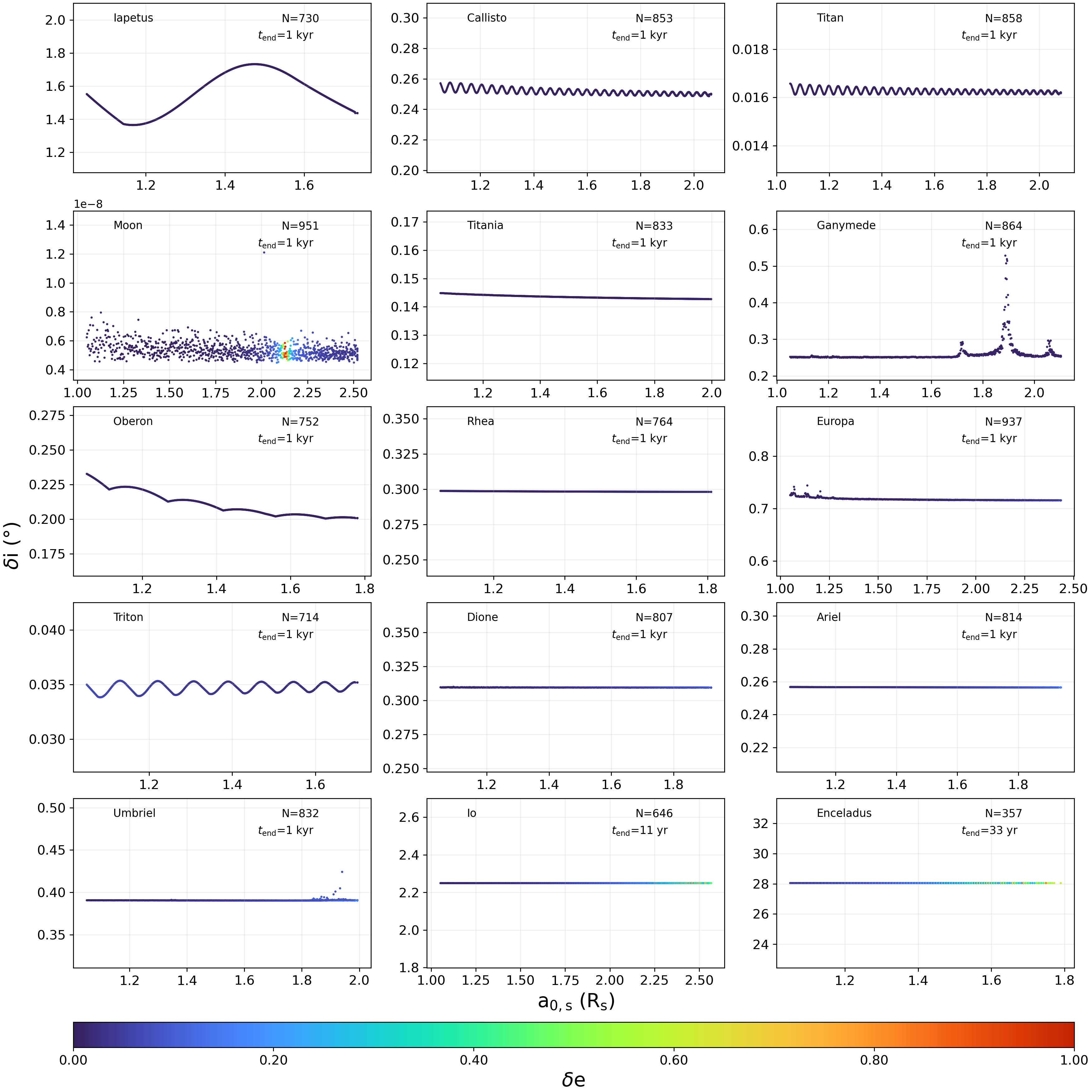}
    \caption{\new{Numerical simulation outcomes for rings around moons. $\delta e$ and $\delta i$ represent the maximum amplitude of the excitation of the ring particles' eccentricity and tilt, respectively. $N$ denotes the number of surviving particles after a simulation time of $t_\mathrm{end}$}. }
    \label{fig:resonances-i-giant}
\end{figure*}
%FFFFFFFFFFFFFFFFFFFFFFFFFFFFFFFFFF
%FFFFFFFFFFFFFFFFFFFFFFFFFFFFFFFFFF
Studying the dynamics of particles within a ring system can lead to a better understanding of the gravitational environment that encompasses it, potentially revealing regions where the influence of nearby celestial bodies is either more or less prominent. This can help us understand the complex gravitational interactions within the system. Furthermore, this knowledge might offer insights into phenomena such as gravitational resonances that play a pivotal role in shaping the system's morphology. For instance, the Kirkwood gaps in the main asteroid belt, primarily caused by orbital resonances with Jupiter, provide an illustrative example. An orbital resonance occurs when two orbiting bodies, in this case, an asteroid and Jupiter, exert a periodic gravitational influence on each other due to their orbital periods being in a simple ratio (e.g., 2:1, 3:2, etc.). For a moon with numerous sources of perturbation, such as its sibling moons, it is feasible that these features may emerge and persist. \new{These perturbations can have a pronounced effect on the ring, potentially leading to its dispersal or to the confinement of its edges.}

In the pursuit of identifying morphological structures within the rings, we will adopt two approaches. Firstly, we will analyse the orbital widening of the test particles that comprise them. Secondly, we will examine the rings as a whole, assessing the emergence of substructures within them. With respect to the first approach, we define the maximum orbital width of each particle, $\Delta_\mathrm{r,p}$, as the difference between the maximum apoapsis, $Q_\mathrm{i,max}$, and the minimum periapsis, $q_\mathrm{i,min}$, of each particle throughout the total integration time, as expressed in the following equation:

\begin{equation}
    \Delta_\mathrm{r,p} = Q_\mathrm{i,max} - q_\mathrm{i,min}. 
\end{equation}

This indicator reflects how a particle's orbit expands or contracts during its interaction with the environment. Consequently, substantial variations in $\Delta_\mathrm{r,p}$ (measured in units of $\Delta_\mathrm{r}$, the initial ring width) could potentially lead to collisions between particles. These collisions could, in turn, accelerate the decay of their orbits, contribute to the formation of moonlets (see e.g. \citealt{Crida2012,Cuk2020}), or result in the loss of particles in specific regions. The results are summarised in Fig.~\ref{fig:dQq}.

According to this figure, Mimas, Miranda, and Proteus undergo significant changes in this parameter, thus aligning with the results obtained from evaluating the MEGNO parameter (Fig.~\ref{fig:megno}). There is also a noticeable orbital widening of the particles in the rings of Io, Europa, Enceladus, Dione, Ariel, and Umbriel, which, as shown in Fig.~\ref{fig:orbdom}, are situated in the uppermost positions close to the region of orbital instability. Furthermore, the presence of narrow regions in the rings of the Moon, Callisto, and Titan, where the orbital width is high, suggests the emergence of gaps due to resonant interactions with the environment.

We also inspected morphological structures in the rings through variations in the semi-major axis, eccentricity, and inclination of the particles. Fig. \ref{fig:resonances-e-giant} and Fig. \ref{fig:resonances-i-giant} illustrates the collective evolution of CSRs, which are organised to show the least dispersed data at the top and the most dispersed at the bottom, with respect to their eccentricity. In Fig. \ref{fig:resonances-e-giant}, the excitation of particle eccentricity relative to their initial semi-major axis is shown. Generally, it is noticeable that there are three modes among the data dispersion: firstly, rings whose dispersion is minimal, such as in the case of Iapetus; secondly, those whose orbits closest to the satellite experience greater excitation, as with the rings of Oberon and Triton; and thirdly, those where the outer regions are the most disturbed, as in the rest of the cases studied. However, it is important to mention that the rings of the Moon, Ganymede, and Oberon exhibit $\delta i<0.01$, suggesting a general stability of the rings.

\begin{table*}[]
\centering
\begin{tabular}{@{}llllr@{}}
\toprule
\multicolumn{1}{c}{Satellite} & \multicolumn{4}{c}{Experiments}                                                             \\ \midrule
\multicolumn{1}{c}{} & \multicolumn{1}{c}{LTI}  & \multicolumn{1}{c}{$Y$}  & \multicolumn{1}{c}{$\delta a$, $\delta e$, $\delta 1$} & Surviving particles \\ \midrule
Moon                          & \cellcolor[HTML]{329A9D} & \cellcolor[HTML]{329A9D} & Cold ring + structure (gap)   & 95 \% \\
Io                            & \cellcolor[HTML]{FFCC67} & \cellcolor[HTML]{FFCC67} & Hot ring                      & 65\%  \\
Europa                        & \cellcolor[HTML]{FFCC67} & \cellcolor[HTML]{329A9D} & Cold ring  + structure (gaps) & 94\%  \\
Ganymede                      & \cellcolor[HTML]{329A9D} & \cellcolor[HTML]{329A9D} & Cold ring + structure (gaps)  & 86\%  \\
Callisto             & \cellcolor[HTML]{329A9D} & \cellcolor[HTML]{329A9D} & Cold ring + structure (gaps and waves)                 & 85\%                \\
Mimas                         & \cellcolor[HTML]{FD6864} & \cellcolor[HTML]{FD6864} & Depleted                      & 0\%   \\
Enceladus                     & \cellcolor[HTML]{329A9D} & \cellcolor[HTML]{FFCC67} & Hot ring                      & 35\%  \\
Dione                         & \cellcolor[HTML]{329A9D} & \cellcolor[HTML]{329A9D} & Warm ring                     & 80\%  \\
Rhea                          & \cellcolor[HTML]{329A9D} & \cellcolor[HTML]{329A9D} & Cold ring                     & 76\%  \\
Titan                & \cellcolor[HTML]{329A9D} & \cellcolor[HTML]{329A9D} & Cold ring + structure (gaps and waves)                 & 86\%                \\
Iapetus                       & \cellcolor[HTML]{FFCC67} & \cellcolor[HTML]{329A9D} & Cold ring + structure (waves) & 73\%  \\
Miranda                       & \cellcolor[HTML]{FD6864} & \cellcolor[HTML]{FD6864} & Depleted                      & 0\%   \\
Ariel                         & \cellcolor[HTML]{329A9D} & \cellcolor[HTML]{329A9D} & Warm ring                     & 81\%  \\
Umbriel                       & \cellcolor[HTML]{329A9D} & \cellcolor[HTML]{329A9D} & Warm ring + structure (gap)   & 83\%  \\
Titania                       & \cellcolor[HTML]{329A9D} & \cellcolor[HTML]{329A9D} & Cold ring + structure (gaps)  & 83\%  \\
Oberon                        & \cellcolor[HTML]{FD6864} & \cellcolor[HTML]{329A9D} & Cold ring + structure (waves) & 75\%  \\
Proteus                       & \cellcolor[HTML]{FD6864} & \cellcolor[HTML]{FD6864} & Depleted                      & 0\%   \\
Triton                        & \cellcolor[HTML]{FD6864} & \cellcolor[HTML]{329A9D} & Cold ring + structure (waves) & 71\%  \\ \bottomrule
\end{tabular}
\caption{Table summarizing the results of the simulations. In the first two columns, the colors indicate the following: blue for regular behavior, red for irregular behavior, and yellow for size reduction for long term integration (LTI) and Megno (Y). The third column describes the ring morphology, where ``cold'', ``warm', and ``hot'' classify eccentricities as $e\leq0.1$, $0.1<e<0.2$, and $e>0.2$, respectively. The final column shows the number of surviving particles.}
\label{tab:summary_exp}
\end{table*}

Fig. \ref{fig:resonances-e-giant}  and Fig. \ref{fig:resonances-i-giant} also shows that certain moons display evident perturbation patterns that can be seen as peaks in the data. This is the case for Callisto, Titan, the Moon, Titania, and Ganymede, suggesting regions of instability in the ring where particles could collide with each other, decay, and/or form gaps. These gaps would be located in different regions of the rings: in the inner part for Titania and Ganymede, in the central part for Callisto and Titan, and in the outermost region of the Moon's rings.

For Io and Enceladus, we observed that about 50\% of the ring particles vanished within just a decade. This significant loss can be attributed to factors such as close encounters with the moon leading to ejections or collisions, and the influence of Jupiter's strong gravitational field, amplified by the perturbations from its other moons. For both moons, these patterns of particle dispersion were observed over simulation periods of 33 and 11 years, respectively.

Fig.~\ref{fig:resonances-i-giant} illustrates the maximal excitation of the inclination for each satellite ring system, and they are presented in the same order as on the left panel. In general, the excitations in inclination are relatively low, except for Io where $\delta i\sim2.5^\circ$, and Enceladus, whose values reach 28° for the surviving particles. For the rest of the rings, these variations are less than 1.5 degrees on average, with Iapetus' rings being the most affected ones.

\new{Also, several of the structures exhibit a wave-like behaviour (e.g., Iapetus, Callisto, Titan, Oberon, and Triton), while others, regardless of the scale, appear quite flat (e.g. Titnia, Rhea, Dione, Ariel, and others), with slight dispersion in the inclinations that mostly do not exceed a hundredth of a degree.}

Additionally, \new{ On one hand, several trends can be seen in the data. Firstly, oscillatory patterns are evident in the rings of Iapetus, Callisto, Titan, Oberon, and Triton, which could be interpreted as density waves traversing the rings. However, these patterns may be dumped by the mutual gravitational interaction between particles with mass, which are not considered in this work (e.g., \citealt{Lehmann2019}). On the other hand, while others, regardless of the scale, appear quite flat (e.g. Titania, Rhea, Dione, Ariel, and others), with slight dispersion in the inclinations that mostly do not exceed a hundredth of a degree.}

Ganymede, Europa, and Umbriel show areas where particles experience great excitation in their inclination compared to their companions. For example, the triple peak shown by Ganymede does not have the same location as the eccentricity peak present in the respective figure of the left panel, so the rings of this moon could be very rich in substructures. The other moons showed a flatter behaviour and did not exhibit major features in the studied orbital elements.

Moons such as Io, Callisto, Dione, Rhea, Ariel, Oberon, and Titania display a discernible relationship between inclination variations and the initial semi-major axis $a_\mathrm{0,s}$. Specifically, within Io, $
\delta i$ appear more pronounced for particles positioned at greater distances. Conversely, for Callisto and Titania, there's a decrease in inclination variation relative to their proximity to the moon. Hence, both Io and Enceladus might retain ringlets in closer orbits, but additional factors (radiation, magnetic fields, higher-order gravitational effects) might challenge their stability.

Iapetus exhibits a constrained dispersion, primarily at low $\delta e$ values, implying minimal eccentricity perturbation for particles in its vicinity. The plot's scale underlines the subtle nature of these eccentricity shifts, suggesting particles around Iapetus exist in a relatively stable environment where rings could thrive over extended time-scales, despite its slight variation in inclination of about $1.5^\circ$.

Significant inclination variations of particles are evident in specific scenarios. The case of Iapetus stands out as a compelling one. Positioned approximately 60 planetary radii from Saturn, with Titan, its nearest neighbour, 20 planetary radii away, Iapetus could be perceived as relatively isolated. In spite of this, Iapetus' ring system's shows non-negligible inclination changes, while the eccentricity variations remain trivial. Such unique excitations may be linked to Iapetus' marked inclinations to multiple reference planes: $17.28^\circ$ relative to the ecliptic, $15.47^\circ$ to Saturn's equator, and $8.13^\circ$ towards the Laplace plane\footnote{The Laplace invariant plane is the plane defined by the total angular momentum of a N-body system, in this case, the system formed by the planet and its moons.}. These prominent inclinations might induce the activation of $\nu$-type secular resonances in inclination, which involve interactions between the precession frequencies of orbits' nodes or pericenters and can potentially alter the trajectories of ring particles. A detailed analysis of these $\nu$-type resonances and their impact on ring particles will be the subject of future work. \vv{Moreover, as discussed in sec.~\ref{subs:environment}, we found that the eccentricity peaks appear when only the planet is included in the simulations, that is, without the influence of sibling moons. On the other hand, the inclusion of nearby massive satellites generates the wavy pattern observed in inclination, a parameter that is not significantly affected by the presence of the planet alone.}

\new{In the case of the Earth's Moon, the peaks in its rings are caused by the Earth itself. Experiments including or excluding the Sun showed no significant changes in the ring system, indicating that the Sun's dynamic influence is negligible due to its distance and the ring's extent.}

\new{Finally, a summary of the experiments and their outcomes is presented in Tab.\ref{tab:summary_exp}. In the first two columns, the colors indicate for long-term integration (LTI) and Megno ($Y$), the following: blue for regular behavior, red for irregular behavior, and yellow for size reduction (i.e. depletion of inner or outer particles in the ring). The third column describes the ring morphology, where ``cold'', ``warm'', and ``hot'' classify eccentricities as $e\leq0.1$, $0.1<e<0.2$, and $e\geq 0.2$, respectively. The final column shows the number of surviving particles. Notably, these results align perfectly with Fig. \ref{fig:orbdom}, demonstrating that the particles most affected by their environment are located at the top of the graph, while the most stable ones are at the bottom.}

\section{Discussion and Conclusions}
\label{sec:discussion}

This study explores the complexities of the dynamics and evolution of hypothetical circumsatellital rings (CSRs) within the Solar System. To achieve this, we have carefully selected the most representative moons in the Solar System, based on their mass and size. These moons were equipped with rings positioned within their orbital planes. The ring particles were treated as test particles and randomly distributed using blue noise generation techniques (Sect.~\ref{sec:setup}). All particles were initialised at the same time and different strategies were implemented to decipher the fate of these rings. This involved evaluating the orbital stability of the moons alone (without rings, see Sect. \ref{subs:sat_stab}), integrating representative ring particles (the innermost and outermost) for 1 million years (Sect. \ref{subs:long_term}), analysing the MEGNO parameter for each particle (Sect. \ref{subs:MEGNO}), observing the broadening of the orbital domains for each particle (orbital widening), studying the effect of the gravitational environment on the rings, and investigating the morphological evolution of the whole system (Sect. \ref{subs:environment} and \ref{subs:Morphology}).

As a result of these analyses, we have found that CSRs can demonstrate stability around most of the Solar System's moons within the time frame evaluated. However, it is worth noting that moons such as Mimas, Miranda and Proteus are less likely to have sustained these structures over long periods of time. We have also observed that the constituent particles forming the rings of Io, Europa, Enceladus, Ariel and Umbriel are prone to significant orbital excitation, making the particle system susceptible to collisions and subsequent fading. In addition, resonance patterns have been identified within the rings, suggesting the possible prevalence of structures such as gaps and other phenomena.

The absence of rings around moons is particularly intriguing, given the existence of astrophysical mechanisms that could potentially facilitate their formation. Unfortunately, the processes of ring formation around planets have not yet been directly observed, but the mechanisms underlying their formation have been seen in action. An important event in this respect was the collision of comet Shoemaker-Levy 9 with Jupiter, which occurred after its fragmentation by the tidal forces of the gas giant (see e.g. \citealt{Zahnle1994}). Had the comet hit Jupiter under different impact parameters, it is highly likely that it would have dispersed some or all of its constituent material around the planet (see e.g. \citealt{Hyodo2017}). Among other mechanisms, the most significant for forming rings around moons would be, for example, the grazing collisions of asteroids or comets against these moons, and the continuous emission of particles through cryovolcanic activity.

It is tempting to explain the absence of CSRs by appealing to the complexity of the gravitational environment in which moons are immersed. In this environment, the gravitational field's intensity undergoes significant variations over time, imprinting a chaotic behaviour on particles that would ultimately destabilise the entire ring system. This was the hypothesis we tested in this work. However, we know that circumplanetary rings (CPRs) are structures with unique and complex origins and rapid evolution.

Currently, there is no consensus on the lifespan of CPRs as they are immerse in highly dynamic environments where their durability is influenced by various factors, including particle size, composition, interactions with electric and magnetic fields or solar wind particles, and orbital decay (e.g., \citealt{Durisen2023I}). The last scenario has been studied by \citealt{Sucerquia2022} for constituent particles in CSRs, who found that such rings can be relatively durable at low inclinations; that is, when the rings are face-on with respect to the star. Nevertheless, dusty CPRs are believed to be sustained by the continuous replenishment of dust particles. For instance, interplanetary dust can be captured by the planet's gravity and added to the ring \citep{Colwell1996,Colwell1998}. Additionally, impact ejecta from circumplanetary objects, like moons, can also contribute to the continuous replenishment of dust particles \citep{Burns1999, Kruger1999, Kruger2000, Krivov2002}. In the case of CSRs, cryovolcanism could be the source for replenishing ring particles for their own rings.

The discussion regarding the existence of CSRs has a history. In 2005, the Cassini mission inferred the existence of a ring around Rhea, Saturn's second-largest moon, by observing changes in the flow of electrons trapped by Saturn's magnetic field during its proximity to the moon. This anomaly was attributed to a faint ring or cloud of particles around Rhea, responsible for electron absorption \citep{Kerr2008}. However, later data from the same mission failed to confirm these rings \citep{Tiscareno2010}. Additionally, Iapetus, another of Saturn's large moons, features ``The Iapetus Ridge'', a unique geological formation extending 75\% of the satellite's circumference and reaching up to 20 km in height. Various hypotheses have been proposed for its formation, including endogenous causes like high rotation and exogenous ones such as material deposition from a former massive ring. A consensus on its origin remains elusive.

An endogenous factor potentially causing CSR instability is the strongly prolate shape of some satellites, such as Rhea (\new{refer to Fig.~\ref{fig:triaxial}}). This raises questions about whether a non-spherical object's gravitational field could sustain a ring system. \citet{Lehebel2015} demonstrated that Rhea's gravitational potential allows stable rings only within the equatorial plane, restricting the parameters for long-term stability. Our simulations suggest that Rhea's rings could have short excursions ($\sim 0.3$°) on their inclinations, favouring Rhea potential to host CSRs. \new{However, the referenced study does not account for the influence of nearby moons. Our findings demonstrate that the impact of these moons on the rings is minimal, thus motivating further investigations that incorporate both gravitational effects.}

Our simulations reveal that among the satellites considered, Iapetus has the most stable ring structures. This supports the hypothesis that its prominent equatorial ridge might have originated from the collapse of a dense ring system onto its surface. It is crucial to note that, given Iapetus's significant orbital inclination (see Sect.~\ref{subs:Morphology}), the dynamics of its rings could be markedly influenced by electromagnetic phenomena, in addition to gravitational fields. This inclination leads to Iapetus traversing various regions of Saturn's magnetosphere during its orbit, resulting in a complex interaction between charged particles and the changing magnetic field. However, the extent of the planet's orbit partially mitigates this effect, as evidenced by the considerable Hill radius of the satellite, as illustrated in Fig.~\ref{fig:satellites-prop}. Consequently, the past existence of sub-moons (a moon of a moon, \citealt{Kollmeier2019}) becomes a plausible scenario, which could explain the formation of Iapetus's ridge as a result of the collapse and fragmentation of a sub-moon at its Roche limit, followed by the subsequent decay due to non-gravitational phenomena.

Additionally, it is worth mentioning the dynamic interactions within Jupiter's magnetosphere, particularly concerning its Galilean satellites. Jupiter's extensive magnetic field, encompassing its ring system and the orbits of its four Galilean moons, plays a dynamic role in the interactions within its magnetosphere. These moons, situated near the magnetic equator, serve as both sources and sinks of magnetospheric plasma. Energetic particles, synchronized with Jupiter's rotation and moving at higher speeds than the moons' orbits, collide with their surfaces, causing sputtering and chemical changes via radiolysis \citep{Johnson2004}. This interaction leads to wear and drag on the surfaces and could similarly affect potential low-mass ring particles around the moons, influencing their orbital stability and accelerating decay and erosion processes. Such phenomena, possibly occurring in other moon systems within the Solar System, add complexity to our understanding of these hypothetical celestial bodies.

In the specific case of Io, Jupiter's corotational plasma plays a vital role in considering the potential instability of its hypothetical rings. This plasma's velocity, synchronized with Jupiter's rotation, ranges from a few km/s to high-energy tails of approximately 30-100 km/s \citep{Wilson2002}. Since Jupiter's rotation period exceeds Io's Keplerian velocity, the corotating plasma impacts Io with high-energy ions, potentially inducing instability in any rings Io might possess. This interaction emphasises the need to understand ejected particle dynamics, atmospheric sputtering, and charge exchange. It also suggests a mechanism for the blurring or dispersion of planetary rings not only in Jupiter but in other satellite systems with similar magnetic and orbital configurations. This highlights the intricate relationship between magnetic fields, orbital dynamics, and ring stability, with broader implications for understanding ring systems in our Solar System and beyond.

\new{We have not investigated in depth the reason why the particles in the hypothetical ring of Triton, despite the large mass of the moon, exhibit one of the largest eccentricity and inclination excursions. As pointed out in Sec.~\ref{subs:long_term}, we speculate that, since this is the only regular moon in our sample that is in a retrograde orbit, this might be one of the main reasons for those instabilities. A more in depth theoretical investigation of the case of Triton, and perhaps proper numerical simulations could be ran to confirm our hypothesis or reveal other  reasons. However, since our preliminary results here, reveals that the probability of a ring around this moon is very low, we do not consider to delve more in this particular case.}

In this study, we have delved into the dynamics and evolution of hypothetical circumsatellital rings (CSRs) within the Solar System, utilising a representative selection of moons based on their bulk properties. While we have advanced our understanding of these intricate systems, unanswered questions and areas for future investigation remain, for instance, are Rhea's and Iapetus' orbital and surface features a byproduct of a decaying CSR? This work positively supports such a scenario. However, including the effects of planet and moon deformities, particle interaction with magnetic fields, and radiative effects is paramount to clarify the nuances of CSRs' dynamics. A continued exploration of these systems will not only enrich our understanding of satellite evolution within our own Solar System, but also offer valuable insights into the dynamical interactions between rings and moons, \new{whose understanding can help us identify these characteristics in distant planetary systems \citep{Zuluaga2022,Veenstra2024}}.

\section*{Acknowledgements}
MS acknowledges support from ANID (Agencia Nacional de Investigación y Desarrollo) through FONDECYT postdoctoral 3210605. MS, JC \& MM thank ANID - Millennium Science Initiative Program $-$ NCN19\_171. \vv{MM acknowledges financial support from FONDECYT Regular 1241818.} This project received funding from the European Research Council (ERC) under the European Union Horizon Europe programme (grant agreement No. 101042275, project Stellar-MADE). This research has made use of NASA's Astrophysics Data System Bibliographic Services, a modified A\&A bibliography style file for the preprint version of the article (\href{https://github.com/yangcht/AA-bibstyle-with-hyperlink}
 {https://github.com/yangcht/AA-bibstyle-with-hyperlink})

\section*{Data Availability}
The data underlying this article was stored in the  {\tt SimulationArchive} format \citep{reboundsa} to ensure full reproducibility, and it will be shared on reasonable request to the corresponding author. 
%\appendix
\bibliographystyle{aa_urs}
\bibliography{references.bib}
\appendix

\section{Orbital and physical parameters}
%FFFFFFFFFFFFFFFFFFFFFFFFFFFFFFFFFF
%FFFFFFFFFFFFFFFFFFFFFFFFFFFFFFFFFF
\begin{figure*}
\centering
    \includegraphics[width=0.7\textwidth]{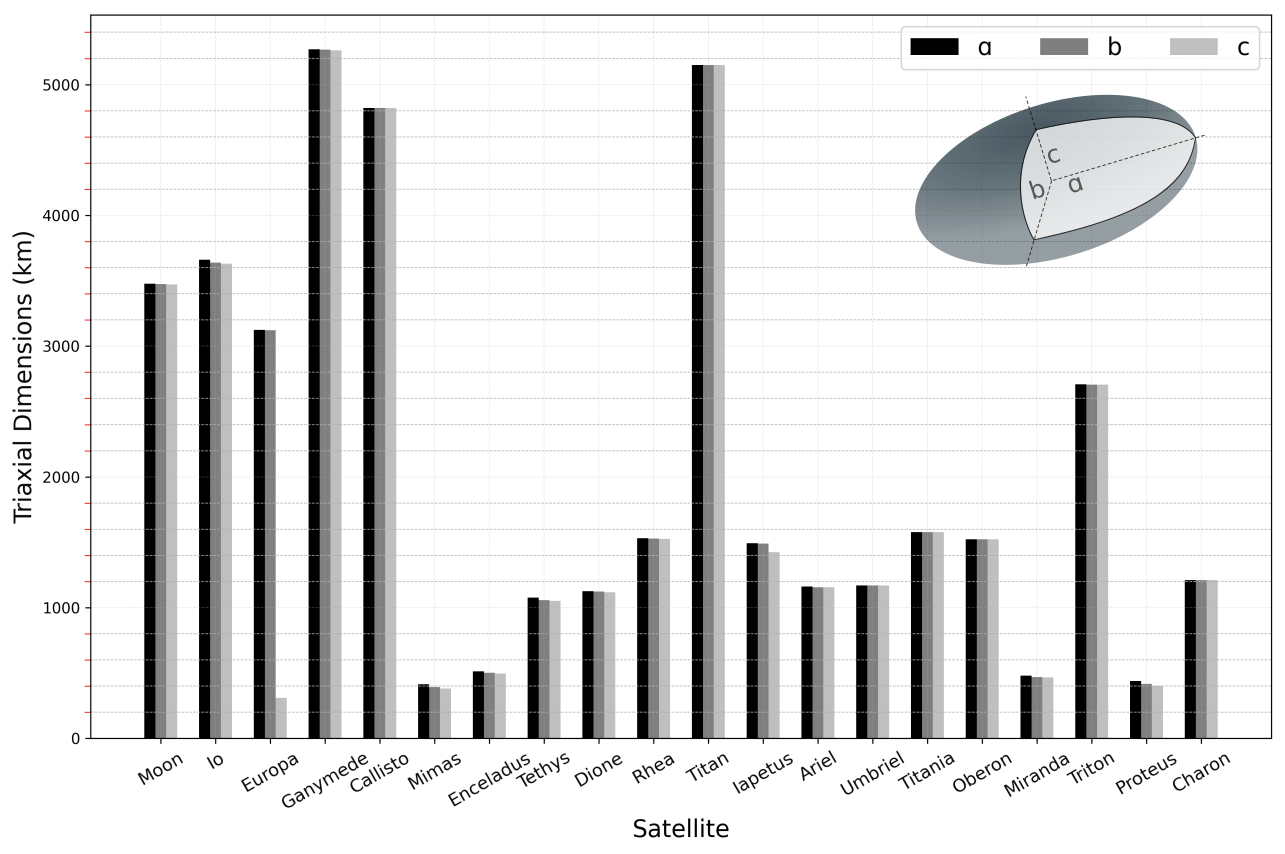}
    \caption{Triaxial dimensions of moons with masses greater than \(10^{19}\) kg. The bars represent the principal axes \(a\), \(b\), and \(c\) for each moon, showcasing their deviation from a perfect sphere.}
    \label{fig:triaxial}
\end{figure*}
%FFFFFFFFFFFFFFFFFFFFFFFFFFFFFFFFFF
%FFFFFFFFFFFFFFFFFFFFFFFFFFFFFFFFFF
\new{Table \ref{tab:physels} and Fig. \ref{fig:triaxial} summarise the physical and orbital parameters of the satellites used in this work. The initial conditions for the planets and moons presented in Table \ref{tab:physels} were generated using the ephemeris service of NASA's Horizons program\footnote{\url{https://ssd.jpl.nasa.gov/horizons/app.html\#/}}, while the physical parameters of Fig. \ref{fig:triaxial} were obtained from the \textit{Physical Data for Solar System Planets and Satellites} compiled by Wm. Robert Johnston\footnote{\url{https://www.johnstonsarchive.net/astro/solar_system_phys_data.html}}.}

\begin{table*}[]
\small
\centering
\begin{tabular}{lccccccccr}
\hline
Body & Type & $a$ (au) & e & $i$ (°) & $\Omega$ (°) & $\omega$ (°) & $M$ (°) & Mass (kg) & Radius (km) \\
\hline
Neptune & Planet & - & - & - & - & - & - & 1.0241$\times 10^{26}$ & 24760 \\
Proteus & Moon & 0.0007863 & 9.77$\times 10^-5$ & 0.5075 & 0.8513 & 5.184 & 0.8779 & 3.8700$\times 10^{19}$ & 208 \\
Triton & Moon & 0.002371 & 0.0002623 & 2.256 & -2.417 & 1.171 & 3.166 & 2.1400$\times 10^{22}$ & 1353 \\
Uranus & Planet & - & - & - & - & - & - & 8.6810$\times 10^{25}$ & 25560 \\
Miranda & Moon & 0.0008681 & 0.001255 & 1.78 & 2.946 & 5.949 & 1.909 & 6.4700$\times 10^{19}$ & 235.8 \\
Oberon & Moon & 0.0039 & 0.000643 & 1.709 & 2.927 & 3.471 & 1.369 & 3.0800$\times 10^{21}$ & 761.4 \\
Titania & Moon & 0.002916 & 0.002244 & 1.706 & 2.926 & 3.945 & 4.987 & 3.4000$\times 10^{21}$ & 788.9 \\
Umbriel & Moon & 0.001778 & 0.003509 & 1.705 & 2.927 & 0.6965 & 5.832 & 1.2800$\times 10^{21}$ & 584.7 \\
Ariel & Moon & 0.001276 & 0.0005182 & 1.705 & 2.926 & 2.958 & 0.07842 & 1.2500$\times 10^{21}$ & 578.9 \\
Saturn & Planet & - & - & - & - & - & - & 5.6834$\times 10^{26}$ & 60270 \\
Iapetus & Moon & 0.0238 & 0.02841 & 0.2981 & 2.423 & 4.044 & 5.278 & 1.8100$\times 10^{21}$ & 734.3 \\
Titan & Moon & 0.008168 & 0.02871 & 0.4835 & 2.951 & 3.067 & 1.621 & 1.3500$\times 10^{23}$ & 2575 \\
Rhea & Moon & 0.003525 & 0.0008698 & 0.4889 & 2.971 & 3.429 & 1.423 & 2.3100$\times 10^{21}$ & 763.5 \\
Dione & Moon & 0.002524 & 0.002243 & 0.4898 & 2.958 & 2.471 & 1.461 & 1.1000$\times 10^{21}$ & 561.4 \\
Enceladus & Moon & 0.001594 & 0.00349 & 0.4895 & 2.959 & 5.033 & 3.885 & 1.0800$\times 10^{20}$ & 252.1 \\
Mimas & Moon & 0.001244 & 0.02219 & 0.484 & 3.016 & 2.457 & 5.876 & 3.7500$\times 10^{19}$ & 198.2 \\
Jupiter & Planet & - & - & - & - & - & - & 1.8982$\times 10^{27}$ & 71490 \\
Callisto & Moon & 0.01258 & 0.007333 & 0.0343 & -0.4016 & 0.5385 & 5.077 & 1.0800$\times 10^{23}$ & 2410 \\
Ganymede & Moon & 0.007157 & 0.002043 & 0.04048 & -0.3465 & 5.985 & 2.979 & 1.4800$\times 10^{23}$ & 2631 \\
Europa & Moon & 0.004487 & 0.009524 & 0.04396 & -0.5335 & 2.355 & 1.011 & 4.8000$\times 10^{22}$ & 1561 \\
Io & Moon & 0.002821 & 0.003984 & 0.03913 & -0.4037 & 5.533 & 1.842 & 8.9300$\times 10^{22}$ & 1821 \\
Earth & Planet & - & - & - & - & - & - & 5.9724$\times 10^{24}$ & 6371 \\
Moon & Moon & 0.002546 & 0.04678 & 0.09025 & 1.087 & 2.774 & 0.7622 & 7.3500$\times 10^{22}$ & 1740 \\
\hline
\end{tabular}
\caption{Orbital elements (semi-major axis ($a$), eccentricity ($e$), inclination ($i$), longitude of ascending node ($\Omega$), argument of periapsis ($\omega$), and mean anomaly ($M$)), masses, and radii of planets and their moons for the epoch 2021-10-11 15:33.}
\label{tab:physels}
\end{table*}

\end{document}